\documentclass[paper]{JHEP3} 

\JHEPspecialurl{http://jhep.sissa.it/JOURNAL/JHEP3.tar.gz}

\usepackage{epsfig,multicol}
\usepackage{cite}

\usepackage{amsmath}
\usepackage{amssymb}
\preprint{
KUNS-1883\\
KEK-TH-929\\
hep-th/0401038\\}

\title{
Nonperturbative studies of fuzzy spheres
in a matrix model with the Chern-Simons term}
\author{ Takehiro Azuma${}^a$, Subrata Bal${}^a$, 
Keiichi Nagao${}^b$, Jun Nishimura${}^b$ \\
\llap{$^a$}Department of Physics, Kyoto University, 
Kitashirakawa,\\
Kyoto 606-8502, Japan  \\
\llap{$^b$}Institute of Particle and Nuclear Studies, \\ 
High Energy Accelerator Research Organization (KEK),\\
1-1 Oho, Tsukuba 305-0801, Japan  \\
\email{azuma@gauge.scphys.kyoto-u.ac.jp, 
subrata@gauge.scphys.kyoto-u.ac.jp, 
nagao@post.kek.jp, 
jnishi@post.kek.jp}}

\abstract{

Fuzzy spheres appear as classical solutions in a matrix model 
obtained via dimensional reduction of
3-dimensional Yang-Mills theory with the Chern-Simons term.
Well-defined perturbative expansion around these solutions
can be formulated even for finite matrix size, 
and in the case of $k$ coincident fuzzy spheres
it gives rise to a regularized U($k$) gauge theory
on a noncommutative geometry.
Here we study the matrix model nonperturbatively by Monte Carlo simulation.
The system undergoes a first order phase transition as we change the
coefficient ($\alpha$) of the Chern-Simons term.
In the small $\alpha$ phase, the large $N$ properties
of the system 
are qualitatively the same as in the pure Yang-Mills model ($\alpha =0$),
whereas in the large $\alpha$ phase
a single fuzzy sphere emerges dynamically.
Various `multi fuzzy spheres' are observed as meta-stable states,
and we argue in particular that 
the $k$ coincident fuzzy spheres cannot be realized as the true vacuum
in this model even in the large $N$ limit.
We also perform one-loop calculations of various observables 
for arbitrary $k$ including $k=1$. Comparison with our Monte Carlo
data suggests that higher order corrections are suppressed
in the large $N$ limit.

}


\newcommand{\bel}{\begin{equation}\label}

\newcommand{\non}{\nonumber \\}
\newcommand{\n}{\nonumber}
\newcommand {\beq}{\begin{equation}}
\newcommand {\eeq}{\end{equation}}
\newcommand {\beqa}{\begin{eqnarray}}
\newcommand {\eeqa}{\end{eqnarray}}
\newcommand {\bc}{\begin{center}}
\newcommand {\ec}{\end{center}}
\newcommand {\tr}{{\rm tr\,}}

\newcommand {\ee}{\mbox{e}}

\newcommand {\dd}{\mbox{d}}

\def\dag{\dagger}

\def\vs5{\vspace*{5mm}}
\def\vs1{\vspace*{1cm}}
\def\vs2{\vspace*{2cm}}
\def\hs5{\vspace*{5mm}}
\def\hs1{\hspace*{1cm}}
\def\hs2{\hspace*{2cm}}
\def\vs50{\vspace*{50mm}}
\def\vs20{\vspace*{20mm}}

\def\tr{\hbox{tr}}

\begin{document}

\section{Introduction}

Quantization of gravity is one of the most important
problems in particle physics, and lots of attempts have been made so far.
String theory, in particular, enables us to quantize small fluctuations 
of the space-time metric around certain backgrounds, 
but the nonperturbative determination of the background itself is yet 
to be done.
Such an issue may be addressed by studying matrix models, 
which are proposed as nonperturbative formulations of string theory.
For instance, the IKKT matrix model \cite{9612115,9908038}, 
which can be obtained via dimensional reduction of 
10-dimensional super Yang-Mills theory, 
is conjectured to be a nonperturbative definition of 
type IIB superstring theory in ten dimensions.

In this model the space-time is represented by the eigenvalue 
distribution of ten bosonic matrices, and 
if the distribution collapses to a four-dimensional hypersurface,
we may naturally understand the dimensionality of our
space-time as a result of the nonperturbative dynamics of 
superstring theory \cite{Aoki:1998vn}.
From the path-integral point of view,
this phenomenon may be caused
by the phase of the fermion determinant \cite{NV},
and Monte Carlo results support this mechanism \cite{sign}.
In Ref.\ \cite{Nishimura:2001sx} the first evidence for the above
scenario has been obtained by calculating the free energy of
the space-time with various dimensionality
using the Gaussian expansion method up to the 3rd order.
Higher-order calculations \cite{KKKMS}
and the tests of the method itself in simpler models
\cite{Nishimura:2002va}
have strengthened the conclusion considerably.
Another indication of this phenomenon has been
obtained recently from the calculations of the 2-loop effective action 
around fuzzy spheres \cite{0307007}.
See also Refs. \cite{Ambjorn:2000dx,Burda:2000mn,%
Ambjorn:2001xs,exact,Vernizzi:2002mu} for related works.

%

Since the space-time is described by matrices in the IKKT model,
it is generically noncommutative.
If one expands the model around a D-brane background for instance,
one obtains a gauge theory
on noncommutative (NC) geometry \cite{9908141}.
\footnote{
For an earlier work on the connection between matrix models and
NC geometry in the context of toroidal compactification, 
see \cite{CDS}. A mathematical formulation of field theories
on NC geometry is given in Ref.\ \cite{Connes}.
For a 
comprehensive review including recent developments,
see Ref.\ \cite{Szabo:2001kg}.
}
This connection was also understood solely in terms of string 
theory \cite{SW}, which triggered much interest
in the dynamical properties of field theories on NC geometry in general.
It was found that 
various interesting phenomena which do not have the commutative
counterparts occur due to the so-called 
UV/IR mixing effects \cite{Minwalla:1999px}.
%
%
%
%
%
%
%
By considering finite-dimensional matrices,
one may naturally regularize NC field theories.
In particular 
the twisted reduced models \cite{Gonzalez-Arroyo:1982hz},
which appeared in history as an equivalent description 
\cite{Eguchi:1982nm} of large $N$ gauge theories,
can be interpreted as a lattice regularization of field theories
on a noncommutative torus \cite{Ambjorn:1999ts}.
This enables nonperturbative studies of 
various dynamical issues in these theories
by Monte Carlo simulation
\cite{Bietenholz:2002ch,Bietenholz:2002ev,Ambjorn:2002nj}.
%
%
See Ref.\ \cite{Nishimura:2003rj} for a general review on the dynamics of 
matrix models in the context of superstring theory and noncommutative
geometry.

A different, although closely related, type of regularization 
is known as the `fuzzy sphere' \cite{Madore},
which can also be formulated using finite-dimensional matrices.
The UV regularization in this case is introduced by putting an upper bound
on the angular momentum when one expands a function on the sphere
in terms of spherical harmonics. This cutoff procedure is compatible 
with the `star-product' of functions which appears in NC geometry.
Since the regularized theory does not break the continuous symmetries
of the sphere, the well-known problems in lattice field theory
concerning chiral fermions and supersymmetry may be easier to
overcome \cite{Grosse:1995ar}.
The first challenge in this direction is to remove the UV/IR
mixing effects in the `continuum limit'.

Being the simplest
curved compact noncommutative space, 
the fuzzy 2-sphere has been studied
extensively in the literature.
One of the fundamental issues
is the construction of the Dirac operator.
The one proposed in Refs.\ \cite{Grosse:1994ed}
does not have the fermion doubling problem 
\cite{Carow-Watamura:1996wg,balagovi}
and it yields the correct chiral anomaly both
in the global form \cite{chiral_anomaly,non_chi,chiral_anomaly2}
and in the local form \cite{AIN}. 
This Dirac operator, however, breaks chiral symmetry explicitly,
and it may therefore be regarded as an analog of the Wilson fermion 
in lattice gauge theory.
On the other hand, an analog of the overlap Dirac operator
\cite{Neuberger:1997fp}, which satisfies
the Ginsparg-Wilson relation \cite{GinspargWilson}
and hence preserves modified chiral symmetry \cite{ML},
is constructed for the free fermion \cite{balaGW} and
for general gauge configurations \cite{AIN2,nagaolat03}.
\footnote{
The overlap Dirac operator for general gauge backgrounds
has been proposed earlier on a NC torus \cite{Nishimura:2001dq},
and used for a construction of chiral gauge theories.
The correct chiral anomaly in this case has been reproduced
in Ref.\ \cite{Iso:2002jc} for arbitrary even dimensions
by using a topological argument \cite{Fujiwara:2002xh}.
In odd dimensions the analogous Dirac operator is used 
to define a lattice NC Chern-Simons theory \cite{Nishimura:2002hw} 
through the parity anomaly \cite{Bietenholz:2000ca}.
}
In this case the chiral anomaly arises from the measure
and the correct results are reproduced in Refs.\ \cite{AIN2,AIN3,Ydri:2002nt}.
Gauge configurations with non-trivial topology
have been found
for the former Dirac operator 
\cite{non-trivial_config,non_chi,Valtancoli:2001gx,Steinacker:2003sd} 
as well as for the latter \cite{Balachandran:2003ay,AIN3}.
The Seiberg-Witten map \cite{SW} has been constructed
also on the fuzzy 2-sphere \cite{Hayasaka:2002db}.

In string theory fuzzy spheres appear
as classical solutions
in the presence of an external Ramond-Ramond field \cite{Myers:1999ps},
and the low-energy effective theory is given
by NC Yang-Mills theory with the Chern-Simons term \cite{Alekseev:2000fd}
(See also \cite{Alekseev:1999bs}).
Classical solutions of the effective theory and their D-brane
interpretation have been studied in Ref.\ \cite{Hashimoto:2001xy}.
In the matrix-model description of string theory,
fuzzy spheres appear as solutions to the classical equation
of motion if one adds the Chern-Simons term representing the
coupling to the external field \cite{0101102}.
Expanding the bosonic matrices around the classical solution, 
one obtains a NC gauge theory on fuzzy spheres. 
The situation is quite analogous to 
the flat D-brane in matrix models \cite{9908141}
except that the fuzzy sphere can be realized even for finite matrices,
which makes various calculations totally well-defined.
Studying fuzzy spheres
is expected to give us a new insight
into the description of curved space-time in matrix models,
which is vitally important in the context of quantum gravity.
A variety of fuzzy-sphere-like solutions in matrix models
have been studied in Refs.\ \cite{0101102,0103192,0108002,Jatkar:2001uh,%
0207115,0303120,0303196,0307007}. 
Fuzzy spheres are also discussed in a matrix model \cite{BMN}
for the M-theory in the so-called pp-wave background
\cite{Dasgupta:2002hx}.

The stability of the fuzzy sphere against quantum fluctuations 
is a non-trivial important issue, which has been discussed
by perturbative calculations \cite{0101102,Valtancoli:2002rx,%
0303120,0307007}
and by the Gaussian expansion \cite{0303196}.
Since matrix models typically have various fuzzy-sphere-like
solutions, it is important to determine which one 
describes the true vacuum.
Such an issue is related to
the dynamical generation of not only the space-time
but also the gauge group, since 
$k$ coincident fuzzy spheres give rise to a NC gauge theory
with the gauge group of rank $k$.
The unified treatment of the space-time and the gauge group is
one of the advantages of considering noncommutative geometry
or the matrix model formulation of string theory.

In this paper we study
a matrix model obtained via dimensional reduction
of three-dimensional Yang-Mills-Chern-Simons theory,
which is known to have fuzzy spheres as classical solutions.
Unlike previous works we perform fully nonperturbative first-principle
calculations using Monte Carlo simulations.
We find that the single fuzzy sphere is nonperturbatively stable
if the coefficient ($\alpha$) of the Chern-Simons term is sufficiently
large, but it collapses to a `solid ball' if $\alpha$ is smaller than
a critical value.
The transition between the two phases is of first order, 
and we observe a strong hysteresis.
Various `multi fuzzy spheres', which are also classical solutions 
of the model, are observed as meta-stable states,
and we argue in particular that the $k$ coincident fuzzy spheres 
cannot be realized as the true vacuum even in the large $N$ limit.
We also perform one-loop calculations of various observables 
for arbitrary $k$ including $k=1$. Comparison with our Monte Carlo
data suggests that higher order corrections are suppressed
in the large $N$ limit.


The rest of the paper is organized as follows. 
In Section \ref{section:def_model} 
we define the model and review how fuzzy spheres
appear in this model as classical solutions.
In Section \ref{section:YangMills} 
we investigate the phase diagram of the model
and demonstrate the existence of the first order phase transition.
In Section \ref{section:prop_single} 
we discuss the properties of the single fuzzy sphere.
In Section \ref{section:width} 
we study the geometrical structure of the dominant 
configurations in each phase.
In Section \ref{section:multi}
we show how various multi fuzzy spheres
appear as meta-stable states and study their properties.
Section \ref{section:conclusion}
is devoted to a summary and discussions.
In Appendix \ref{heat-bath} we comment on
the algorithm used for our Monte Carlo simulations.
In Appendices \ref{eff_derive} and \ref{one-loop-obs-appendix} 
we give a self-contained
derivation of the one-loop results for 
the effective potential as well as various observables.
In Appendix \ref{instability} we discuss the instability of the
$k$ coincident fuzzy spheres.

\section{The Yang-Mills-Chern-Simons matrix model and the fuzzy spheres}
\label{section:def_model}

The model we study in this paper is defined by the partition function
\cite{0101102}
\beqa
Z &=& \int \dd A \, \ee ^{-S} \ ,
\label{bosonicZ} \\
S &=&  N
\, \tr \left( - \frac{1}{4} 
\, [A_{\mu},A_{\nu}]^{2} + \frac{2}{3}\, i \, \alpha 
\, \epsilon_{\mu \nu \rho} A_{\mu} A_{\nu} A_{\rho}
   \right) , 
\label{verydefinition} 
\eeqa
where $A_\mu$ ($\mu = 1, 2 , 3$) 
are $N\times N$ traceless Hermitian matrices.
The integration measure $\dd A$ is defined by
$ \dd A = \prod_{a=1}^{N^2-1} \prod_{\mu = 1}^{3}
\frac{d A_\mu ^a}{\sqrt{2 \pi}} $,
where $A_\mu^a$ are the coefficients in
the expansion
$A_\mu = 
\sum_{a}
A_\mu ^a \, t^a $
with respect to the SU($N$) generators 
$t^a$ normalized as $\tr (t^a t^b) = \frac{1}{2} \delta ^{ab}$.
The coefficient of the first term in 
(\ref{verydefinition}) is fixed, but this does not
spoil any generality since one can always 
rescale $A_\mu$ to bring the action into the present form.

This model may be regarded as the zero-volume limit
of SU($N$) Yang-Mills theory with the Chern-Simons
term in the three-dimensional Euclidean space, 
and it has the SO$(3)$ rotational symmetry
as well as the SU$(N)$ symmetry
$A_{\mu} \to U A_{\mu} U^{\dag}$, where $U\in {\rm SU}(N)$.
There are some points to note here, though.
The parameter $\alpha$ in (\ref{verydefinition})
is chosen to be real in order 
for fuzzy spheres to be classical solutions of the model.
As a result the Chern-Simons term in the action (\ref{verydefinition})
is real unlike the counterpart in ordinary field theory,
where it is purely imaginary and therefore poses a severe technical problem
in Monte Carlo simulation.
Note also that $\alpha$ may take arbitrary real number 
without breaking any symmetries
unlike the ordinary Chern-Simons theory, where
the coefficient is quantized for the invariance
under topologically nontrivial gauge transformations.
Since the model has the duality $\alpha \mapsto - \alpha$ 
associated with the parity transformation $A_\mu \mapsto - A_\mu$,
we restrict ourselves to $\alpha > 0 $ throughout this paper
without loss of generality.


The pure Yang-Mills model ($\alpha = 0$) and its obvious generalization
to $D$ dimensions with $D$ matrices $A_\mu$ ($\mu=1, \cdots , D$)
have been studied by many authors.
In particular the large $N$ dynamics of the model have been studied
by the $1/D$ expansion and Monte Carlo simulations \cite{9811220}.
The partition function was conjectured \cite{Krauth:1998yu}
and proved \cite{Austing:2001bd} to be finite for $N > D/(D-2)$.
(See Refs. \cite{Krauth:1998xh,Austing:2001pk} for the supersymmetric
case.)
The partition function in the presence of the Chern-Simons term has been
studied analytically for $N=2$ \cite{0309264},
and it turned out to be convergent in the supersymmetric case,
but not in the bosonic case.
It is also proved that
adding a Myers term (the Chern-Simons term in the present case)
does not affect the convergence as far as the original path integral 
converges absolutely \cite{0310170},
which means in particular that the partition function (\ref{bosonicZ})
is convergent for $N\ge 4$.
Note that this statement holds despite the fact that 
the classical action (\ref{verydefinition})
is {\em unbounded from below} for $\alpha \neq 0$.

The classical equation of motion in the present model reads
\begin{eqnarray}
   [A_{\nu}, [A_{\nu}, A_{\mu}]] + i \, \alpha \, \epsilon_{\mu \nu \rho}
   \, [A_{\nu}, A_{\rho}] = 0 \ . 
\label{eom} 
\end{eqnarray}
The simplest type of solutions is given by 
the commutative matrices, which can be brought into the diagonal form
\begin{eqnarray}
    A_{\mu} = \textrm{diag} \, (x^{(1)}_{\mu}, x^{(2)}_{\mu}, \cdots ,
    x^{(N)}_{\mu}) 
\label{diagsolution}
\end{eqnarray}
by an appropriate SU($N$) transformation.
This type of solutions exists also at $\alpha=0$,
and the one-loop effective action around it 
has been calculated in Ref.\ \cite{9811220}.
For $\alpha \neq 0$ the one-loop effective action reads \cite{0101102}
\beq
\Gamma_{\rm 1-loop} = \sum_{i<j} \left[
\log \left\{ \left(x^{(i)}_\mu - x^{(j)}_\mu \right)^2  \right\} 
+ \log \left\{
1 - \frac{4 \alpha^2}{\left(x^{(i)}_\mu - x^{(j)}_\mu
\right)^2}  \right\} 
\right] \ ,
\label{eff_diag}
\eeq
where the second term represents the additional piece for $\alpha \neq 0$.
The resulting attractive force between $x^{(i)}_\mu$
makes their distribution shrink
until the perturbative calculation
becomes no more valid \cite{9811220}.

When $\alpha$ is nonzero, there exists
another type of solutions given by
\begin{eqnarray}
     A_{\mu} = \alpha \, L_{\mu} \ , 
\label{fssolution}
\end{eqnarray}
where $L_\mu$ ($\mu = 1,2,3$) satisfy
\begin{eqnarray}
   [L_{\mu}, L_{\nu}] = i \, \epsilon_{\mu \nu \rho} \, 
   L_{\rho} \ . 
\label{fscom} 
\end{eqnarray}
Since the equation (\ref{fscom}) is nothing but
the SU(2) Lie algebra, each of its $N$-dimensional representation 
yields a classical solution.
We denote the $n$-dimensional irreducible representation 
of the algebra by $L_{\mu} ^{(n)}$,
which can be characterized by the identity
\beq
\sum_\mu  \left( L_{\mu} ^{(n)} \right)^2 
= \frac{1}{4} (n^2 -1 ) \, {\bf 1}_{n} \ .
\eeq

Let us consider the case in which $L_\mu$ itself is
given by the irreducible representation.
The corresponding classical solution 
\beq
A_\mu= \alpha \, L_{\mu} ^{(N)}
\label{config_single}
\eeq
satisfies
\beqa
\sum_{\mu} (A_\mu)^2 &=& 
 R^2 \, {\bf 1}_N \ ,
\label{casimir} \\
R &=& \frac{1}{2} \, \alpha \sqrt{N^{2}-1} \ .
\label{rad_single}
\eeqa
If we neglect for the moment the fact that $A_\mu$ are noncommutative,
eq.\ (\ref{casimir}) implies that all the $N$ eigenvalues of $A_\mu$
are distributed on a sphere centered at the origin
with the radius $R$.
Since $A_\mu$ are noncommutative in reality, this classical solution
is called the {\em fuzzy} 2-sphere (In this paper we refer to it as
the `{\em single} fuzzy sphere').

If we consider a general representation,
the corresponding classical solution can be brought into the form
\begin{eqnarray}
A_\mu = \alpha
\begin{pmatrix}
L_{\mu} ^{(n_1)} & & & \cr 
& L_{\mu} ^{(n_2)} & & \cr
& & \ddots & \cr 
& & & L_{\mu} ^{(n_k)} \cr  
\end{pmatrix}  
\label{general_FS}
\end{eqnarray}
by an appropriate SU($N$) transformation, where
\beq
 \sum_{a=1}^{k} n_a = N \ .
\label{summing_block}
\eeq
This solution satisfies
\begin{eqnarray}
\sum_{\mu} (A_\mu)^2 = 
\begin{pmatrix}
 (r_1)^2 \, {\bf 1}_{n_1} & & & \cr 
&  (r_2)^2 \, {\bf 1}_{n_2} & & \cr
& & \ddots & \cr 
& & & (r_k)^2 \, {\bf 1}_{n_k} \cr  
\end{pmatrix}  \ ,
\label{general_Casimir}
\end{eqnarray}
where 
\beq
r_a =  \frac{1}{2}\,  \alpha \sqrt{(n_a)^2 -1} \ ,
\label{rad_multi}
\eeq
meaning that $n_a$ eigenvalues of $A_\mu$
are distributed on a sphere centered at the origin with the radius $r_a$.
In the $k\ge 2$ case we call the solution (\ref{general_FS})
`{\em multi} fuzzy spheres'.

The particular case of `multi fuzzy spheres' that we will focus on
later corresponds to taking $n_1 = \cdots = n_k \equiv n$, and therefore
$N = n\cdot k$ due to (\ref{summing_block}).
The radius (\ref{rad_multi}) for each sphere becomes equal,
and it is proportional to $1/k$ at large $N$.
Expansion of the model around such a configuration, 
which describes $k$ coincident fuzzy spheres,
gives rise to a NC gauge theory
on the fuzzy sphere with the gauge group of rank $k$ \cite{0101102}.
This statement holds also for $k=1$, which corresponds 
to the single fuzzy sphere.


We study the matrix model (\ref{verydefinition})
by Monte Carlo simulations
using the heat-bath algorithm, which was adopted also in the study of
the pure Yang-Mills model ($\alpha = 0$) in Ref.\ \cite{9811220}.
See Appendix \ref{heat-bath} for more details.

 \section{The phase structure}
 \label{section:YangMills}

 In this Section we study the phase structure of the model
 by calculating fundamental quantities such as
 $\langle S \rangle$ and $\langle \frac{1}{N} \tr (A_{\mu})^{2}
 \rangle$
 by Monte Carlo simulations.
 Here we take the initial configuration to be either
 of the following two.
  \begin{equation}
     A^{(0)}_{\mu} =
\left\{ 
\begin{array}{ll}
 \alpha \, L_\mu ^{(N)}   & \mbox{~~~~~(the single fuzzy sphere start)} \\
   0  & \mbox{~~~~~(the zero start)} 
 \end{array}
 \right.
  \label{zeroinitial} 
  \end{equation}
 Fig.\ \ref{action_cycle} shows the results plotted
 against $\alpha$ for $N=8,16,24$.
 The one-loop results obtained from the perturbative expansion around 
 the single fuzzy sphere are plotted for comparison.


      \FIGURE{
          \epsfig{file=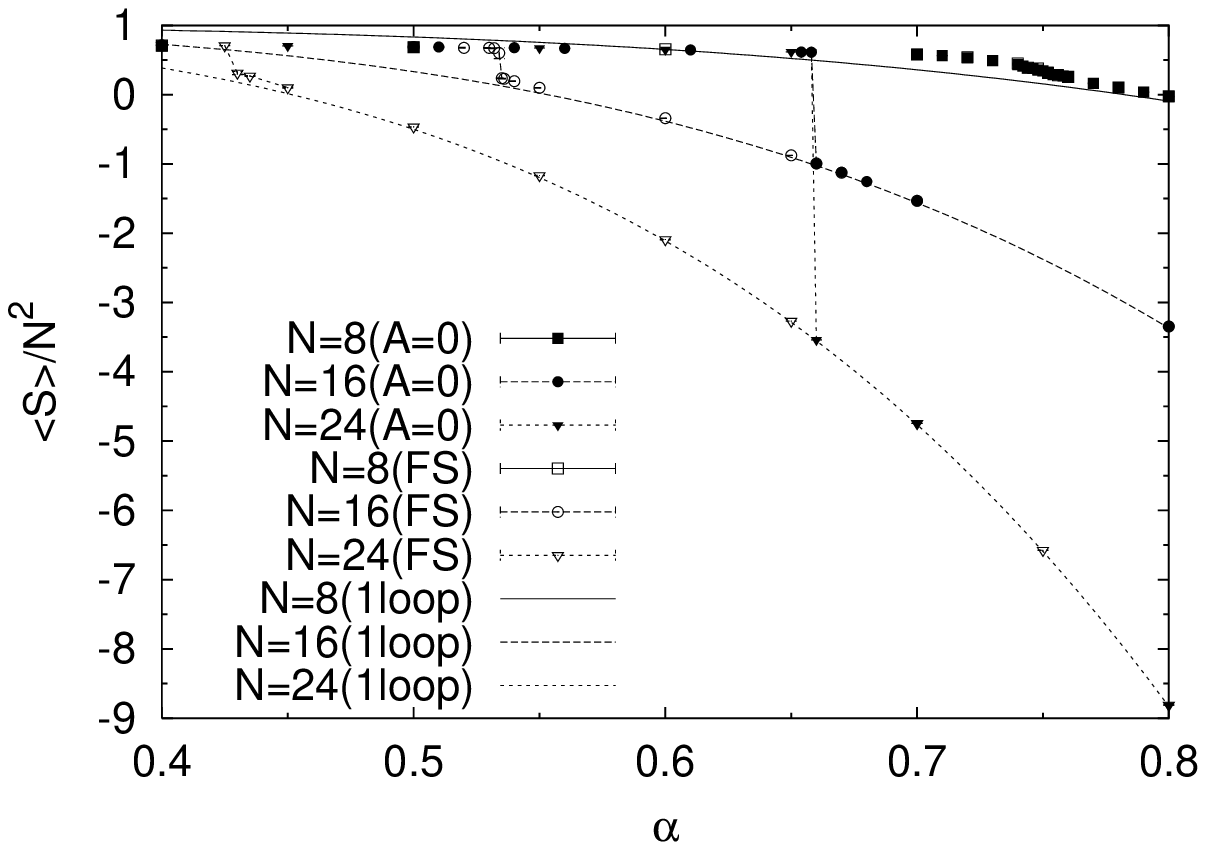,width=7.4cm}
           \epsfig{file=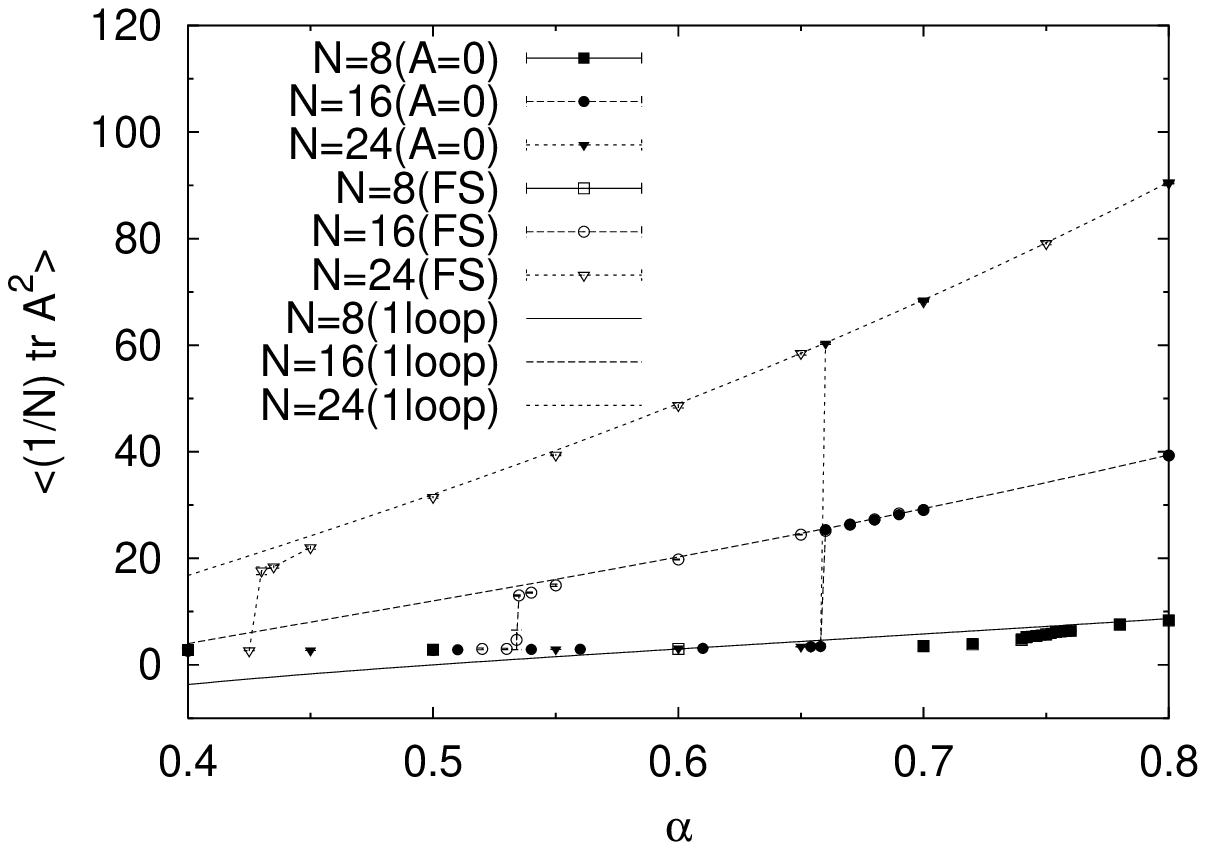,width=7.4cm}
    \caption{The observables $\frac{1}{N^2}\langle S \rangle$ (left)
and $\langle \frac{1}{N} \tr A^{2}_{\mu} \rangle$ (right)
are plotted against  $\alpha$ for $N=8,16,24$.
The open symbols represent the single fuzzy sphere start,
whereas the closed symbols represent the zero start.
We also plot the one-loop results for the single fuzzy sphere
at each $N$ for comparison.
}
  \label{action_cycle}}

The Monte Carlo results depend on the initial configuration
in the intermediate region of $\alpha$ for $N \ge 16$,
and we observe discontinuities at 
\begin{equation}
\alpha = \left\{
\begin{array}{lcll}
\alpha_{\rm cr}^{\rm (l)} &\sim& \frac{2.1}{\sqrt{N}} & 
{\mbox{~~~~~(for the single fuzzy sphere start)}} 
\\
\label{criticalpointzero}
\alpha_{\rm cr}^{\rm (u)} &\sim& 0.66 
& {\mbox{~~~~~(for the zero start)}} \ ,
\end{array} 
\right.
\end{equation}
which we call the lower/upper critical points, respectively.
This clearly demonstrates 
the existence of a first order phase transition.

Note that the large $N$ behavior of the lower critical point is
different from the upper critical point.
If we naively apply the formula (\ref{criticalpointzero})
to $N=8$, we obtain
$\alpha_{\rm cr}^{\rm (l)} = \frac{2.1}{\sqrt{8}}= 0.742$,
which is actually {\em larger} than 
$\alpha_{\rm cr}^{\rm (u)}$.
What happens in reality is that the hysteresis simply disappears
for $N=8$ as seen in Fig.\ \ref{action_cycle}.


In the small $\alpha$ phase
the results do not depend much on $\alpha$.
In particular the large $N$ behavior of each observable is 
qualitatively the same as in the pure Yang-Mills model \cite{9811220}
($\alpha = 0$), {\em i.e.},
$\frac{1}{N^2} \langle S \rangle \sim {\rm O}(1)$ and
$ \left\langle \frac{1}{N} \tr (A_{\mu})^{2} \right\rangle 
\sim {\rm O}(1)$.
We will call this phase the `Yang-Mills phase'.
In the large $\alpha$ phase
the results agree very well 
with the one-loop perturbative calculations around the single
fuzzy sphere,
which we elucidate further in Section \ref{section:prop_single}.
We will call this phase the `fuzzy-sphere phase'.
When we perform simulations with the zero start for
$\alpha > \alpha_{\rm cr}^{\rm (u)}$, 
we observe various
multi fuzzy spheres as meta-stable states in the thermalization
process (See Section \ref{evo}).

\section{Properties of the single fuzzy sphere}
\label{section:prop_single}

In this Section we study 
various properties of the single fuzzy sphere
by Monte Carlo simulations using (\ref{config_single})
as the initial configuration.

\subsection{The lower critical point and the `one-loop dominance'}
\label{section:low_critical}

Here we calculate various observables by Monte Carlo simulations,
and compare the results with the one-loop perturbative calculations.
We determine the lower critical point, and show that our Monte Carlo
data above the critical point agree very well with the one-loop
results.

\FIGURE{\epsfig{file=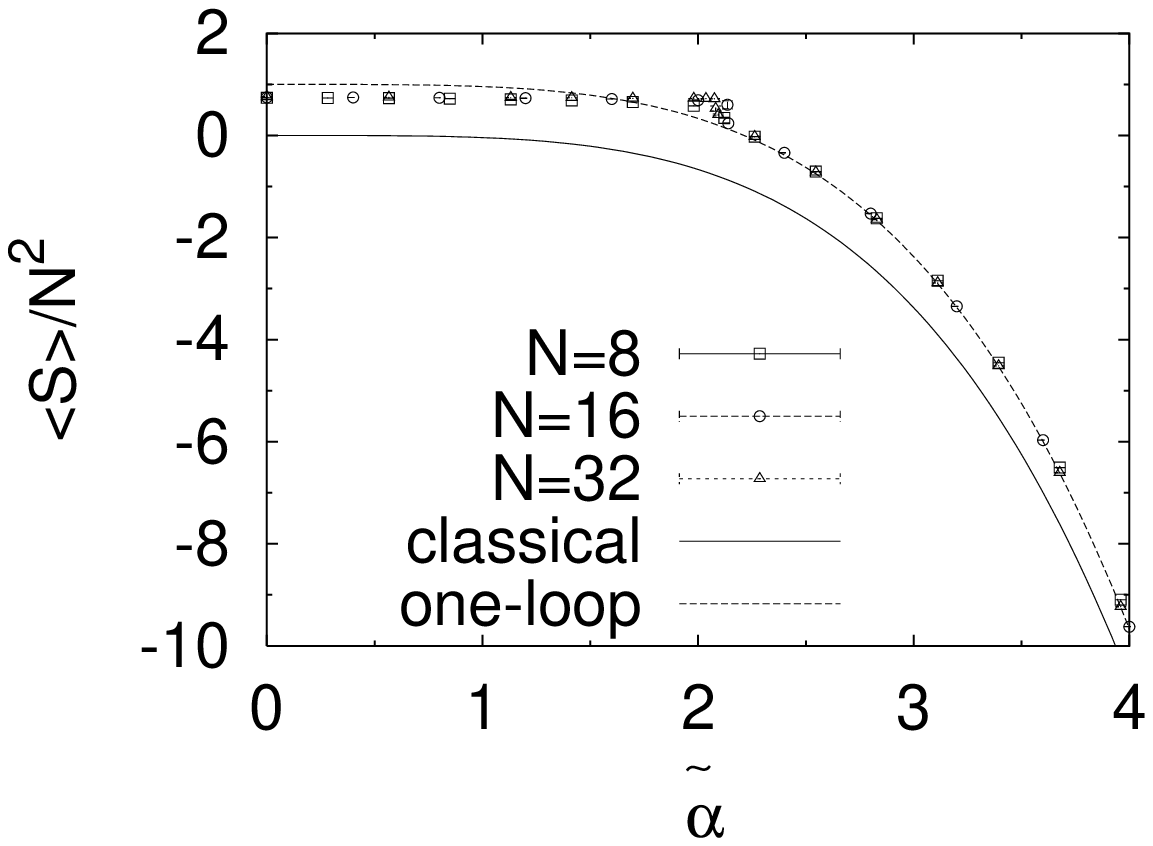,width=7.4cm}
          \epsfig{file=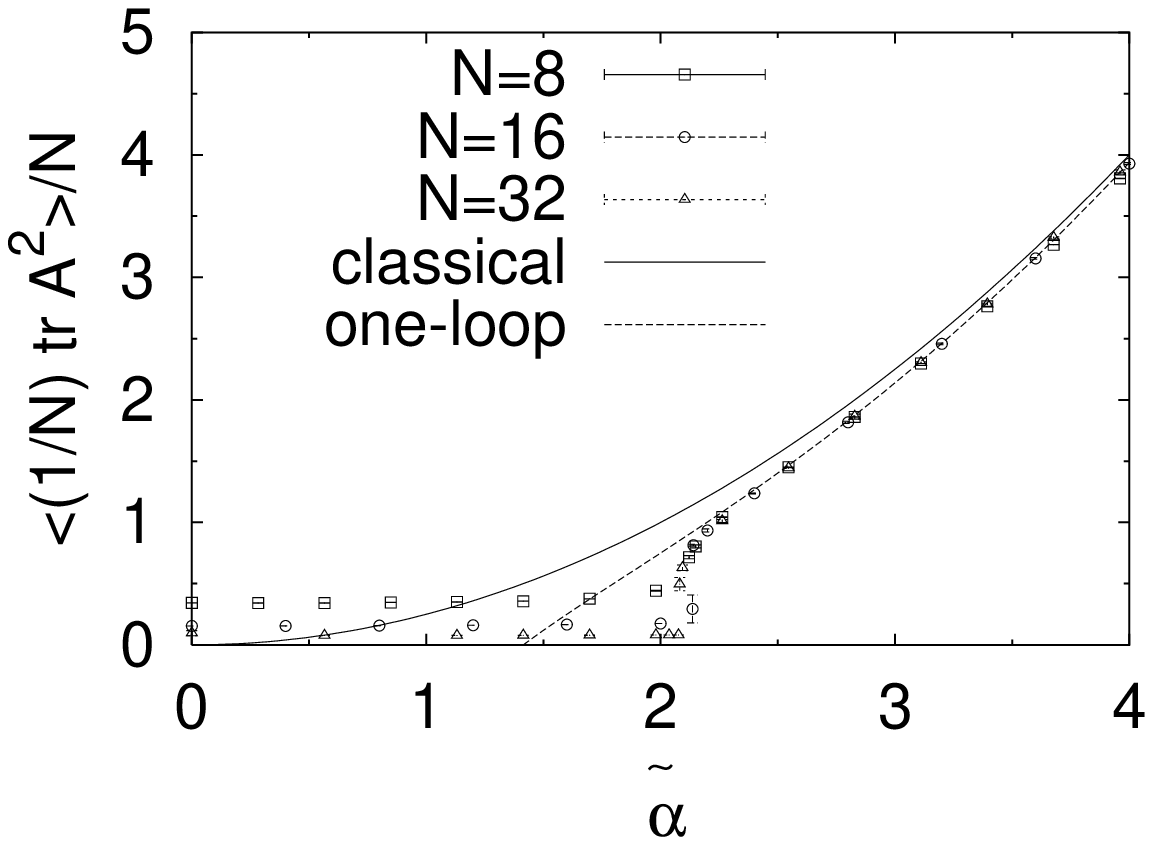,width=7.4cm}
          \epsfig{file=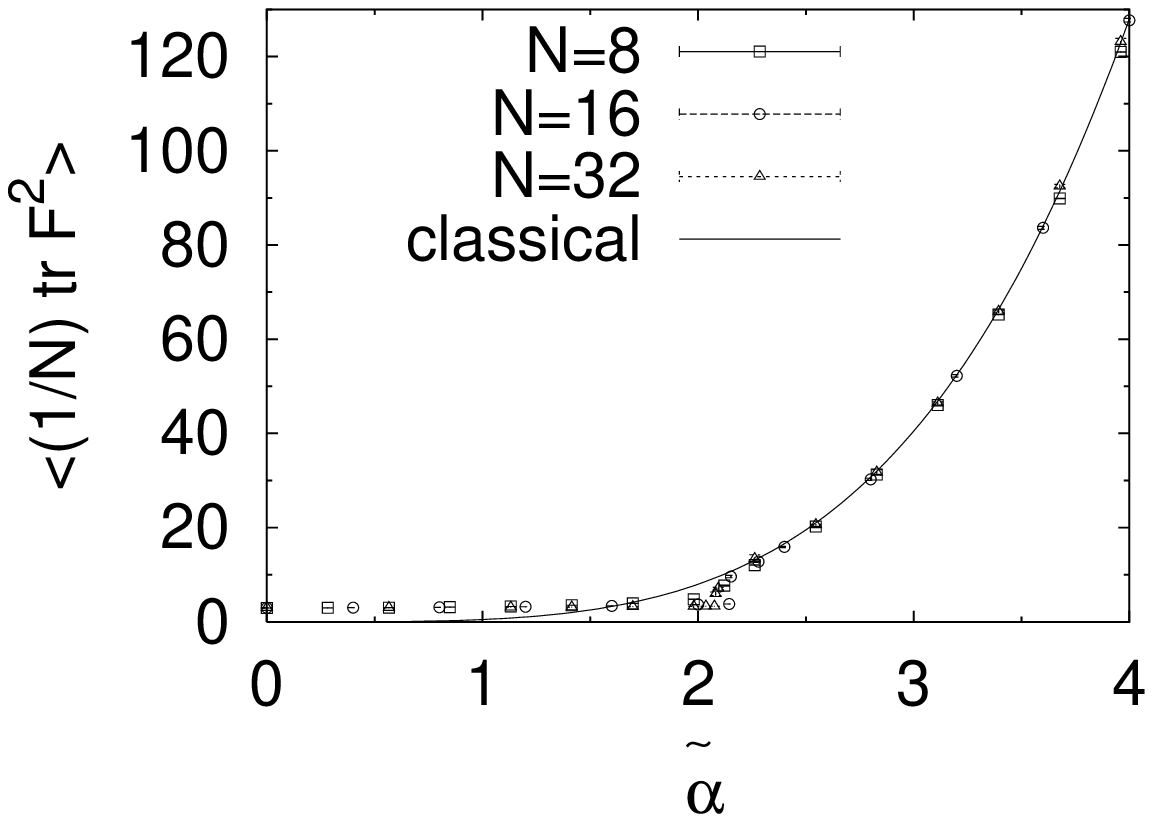,width=7.4cm}
          \epsfig{file=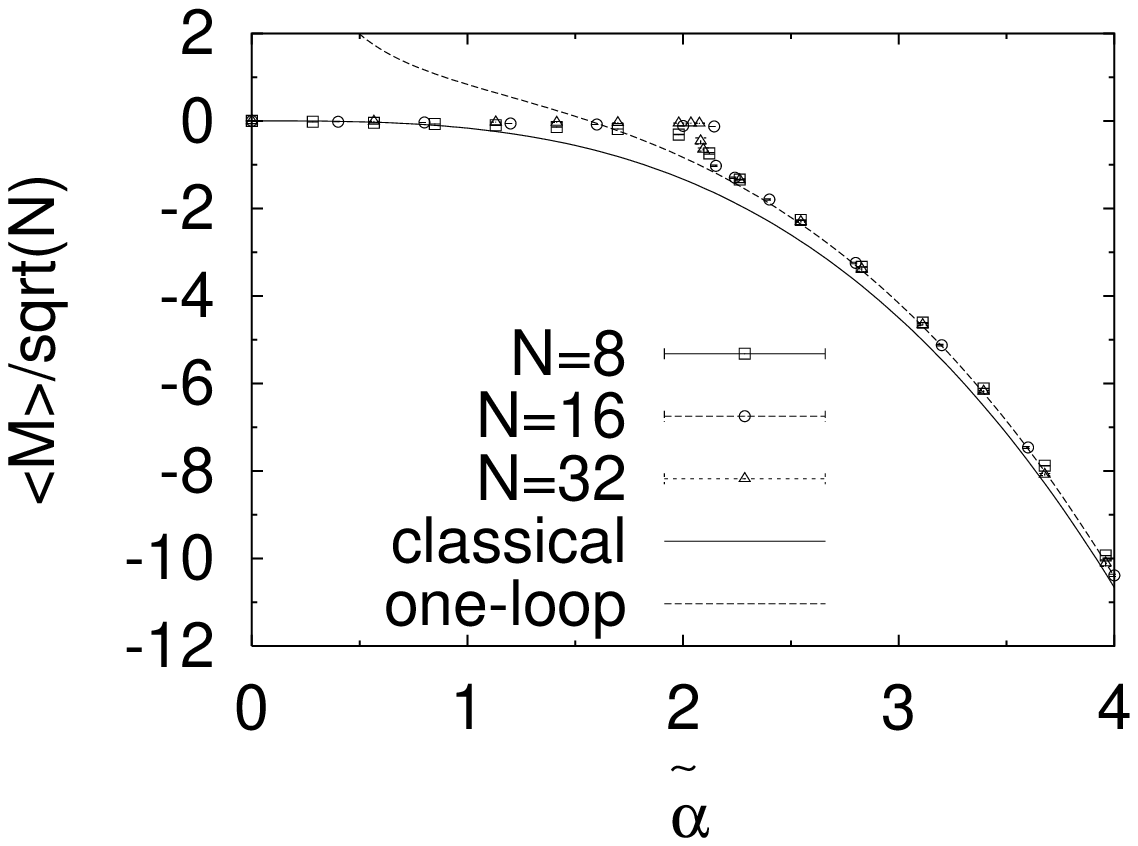,width=7.4cm}
\caption{
Various observables
are plotted against ${\tilde \alpha}= \alpha \sqrt{N}$ 
for $N=8,16,32$
with the single fuzzy sphere start.
The solid lines represent the classical results, whereas the
dashed lines represent the results including the one-loop corrections.
}
\label{miscFS}}

Let us first summarize the results of one-loop calculations.
For brevity we introduce the notation
\beqa
F_{\mu\nu} &=& i \, [A_\mu , A_\nu]  \ , \\
M &=& \frac{2}{3N} \, i   \, 
\epsilon_{\mu \nu \lambda} \, 
\tr  (A_{\mu} A_{\nu} A_{\lambda})  \ ,
\eeqa
so that the action reads $S = N^2 (\frac{1}{4N} \tr F^2 + \alpha M)$.
The leading large $N$ behavior of various observables is given by
\begin{eqnarray}
    \frac{1}{N^{2}}\langle S \rangle_{\rm 1-loop} 
 &\simeq& - \frac{\tilde{\alpha}^{4}}{24} + 1 \ , 
\label{one-loopacto} \\
    \frac{1}{N} \left\langle 
\frac{1}{N} \tr (A_{\mu})^{2} \right\rangle_{\rm 1-loop} 
  &\simeq& \frac{{\tilde \alpha}^{2}}{4} - \frac{1}{{\tilde \alpha}^{2}} \ ,
  \label{one-loopa-sq} \\
   \left\langle \frac{1}{N} \tr (F_{\mu \nu})^{2} \right\rangle_{\rm 1-loop}  
&\simeq& \frac{{\tilde
  \alpha}^{4}}{2} + 0 \ , 
\label{one-loopf-sq} \\
   \frac{1}{\sqrt{N}} \langle M \rangle _{\rm 1-loop} 
&\simeq& - \frac{{\tilde
  \alpha}^{3}}{6} + \frac{1}{{\tilde \alpha}}  \ ,
\label{one-loopcs-a}
\end{eqnarray}
where we have introduced the rescaled parameter
\beq
{\tilde \alpha} = \alpha \sqrt{N} \ .
\label{alpha_tilde}
\eeq
In (\ref{one-loopacto}) $\sim$
(\ref{one-loopcs-a})
the first term represents the classical result,
while the second term represents the one-loop correction
(See Appendix \ref{one-loop-obs-appendix} for derivation).

\FIGURE{\epsfig{file=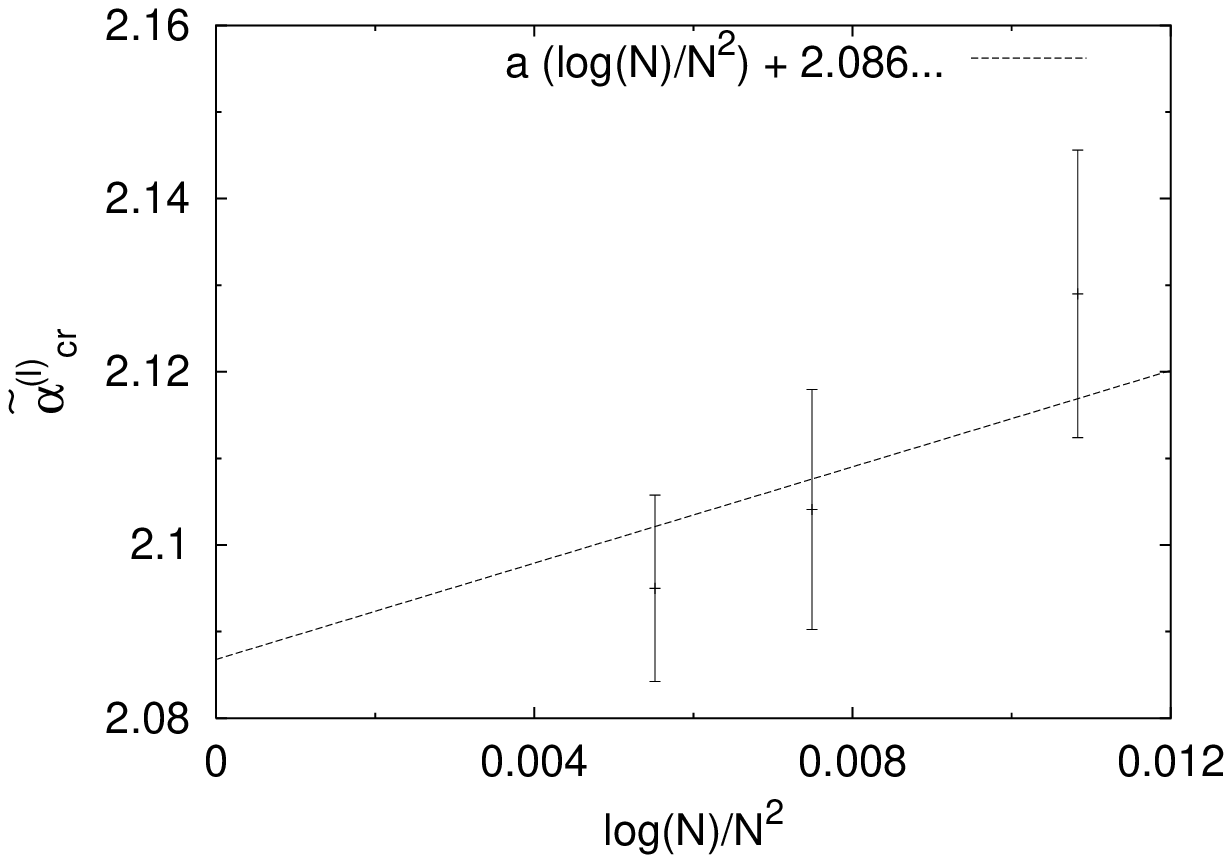,width=7cm}
    \caption{The lower critical point 
$\tilde{\alpha}_{\rm cr}^{\rm{(l)}}$ 
is plotted for $N=16,20,24$. The dashed line represents a fit
to 
$\tilde{\alpha}_{\rm cr}^{\rm (l)} = 
\left( \frac{8}{3} \right)^{3/4} + a  \,  \frac{\log N}{N^2} $,
where $a = 2.781$.}
\label{lower_crit}}

In Fig.\ \ref{miscFS} we plot our Monte Carlo results for 
these quantities against ${\tilde \alpha}$.
All the observables exhibit a discontinuity at 
  \begin{eqnarray}
\tilde{\alpha} = 
\tilde{\alpha} _{\rm cr}^{\rm (l)} \simeq 2.1 \ .
\label{criticalpointfs} 
  \end{eqnarray}
In terms of the original parameter $\alpha$, it
corresponds to the lower critical point 
given in (\ref{criticalpointzero}).

The results above the lower critical point 
scale with $N$, and moreover they agree very well 
with the perturbative results including the one-loop corrections.
For $\langle \frac{1}{N} \tr (F_{\mu \nu})^{2} \rangle$,
the data agree even with the classical result due to
the accidental cancellation of the one-loop correction (\ref{one-loopf-sq}).
These results suggest that the higher order corrections
are suppressed at large $N$ for fixed $\tilde{\alpha}$.

In fact it turns out that the lower critical point (\ref{criticalpointfs})
can be reproduced from the effective potential 
for the scalar mode on the fuzzy sphere \cite{private} as
\footnote{This was informed to us by Denjoe O'Connor
after J.N.\ gave a seminar on the main results of this paper
including (\ref{criticalpointfs})
at Dublin Institute for Advanced Studies.}
\beq
\tilde{\alpha}_{\rm cr}^{\rm (l)} = 
\left( \frac{8}{3} \right)^{3/4} + a  \,  \frac{\log N}{N^2} \ .
\label{pred_eff}
\eeq
In Fig.\ \ref{lower_crit}
we plot the lower critical point $\tilde{\alpha}_{\rm cr}^{\rm (l)}$ 
determined by simulations for $N=8,16,24$,
where the `error bars' represent the range of $\tilde{\alpha}$
for which we observe two-state signals.
Our results are consistent with the large $N$ asymptotic behavior 
(\ref{pred_eff}).

\subsection{Reproducing an exact result}

In this Section we consider an exact result, which can be used 
as a check of Monte Carlo simulations and the `one-loop dominance'
discussed in the previous Section.
Let us note that the expectation value of the observable
\begin{eqnarray}
K = \frac{1}{N} \tr (F_{\mu \nu})^{2} 
+ 3 \alpha M   
\label{sdeqbalance} 
\end{eqnarray}
is given exactly by
\begin{eqnarray}
 \langle K \rangle = 
3\left( 1-\frac{1}{N^{2}} \right)
\equiv  K_{0} 
\label{sde-quantity}
\end{eqnarray}  
for arbitrary $\alpha$ and $N$.
This identity has been studied
also in the $\alpha = 0$ case \cite{9811220}.

\FIGURE{\epsfig{file=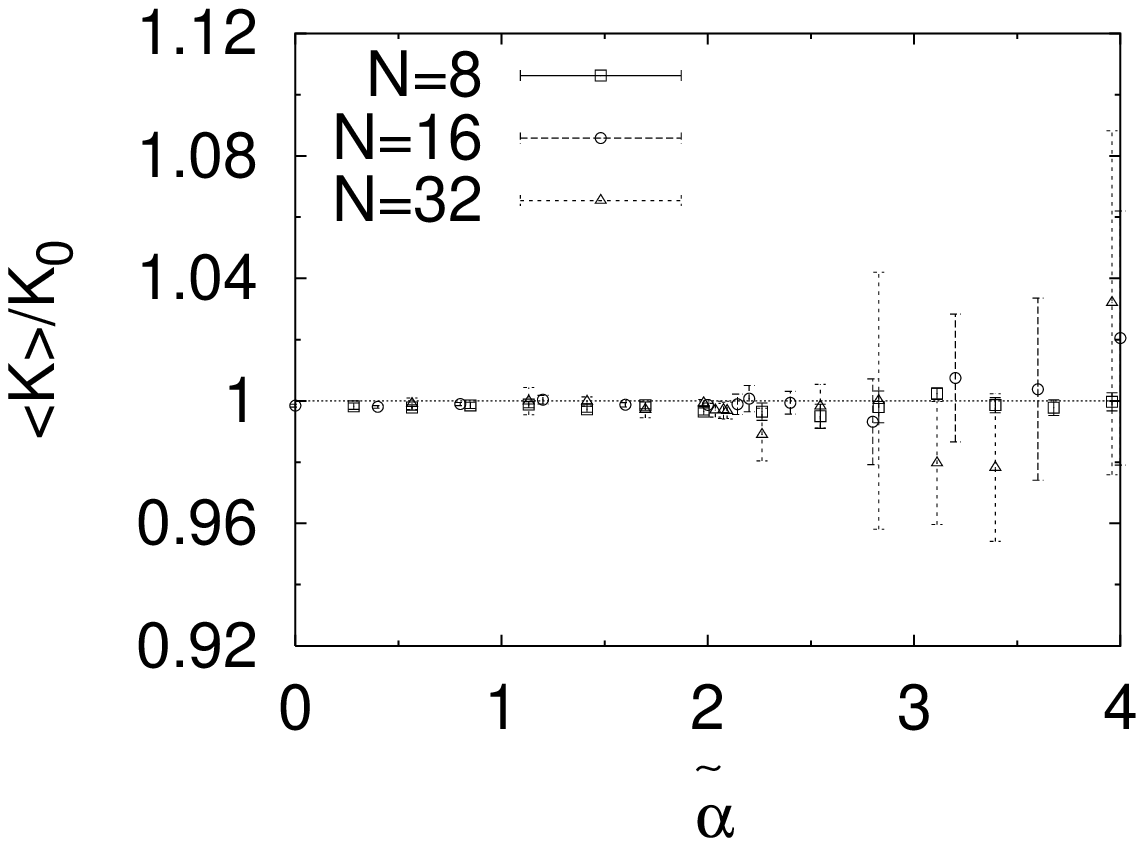,width=7cm}
    \caption{The ratio $\frac{\langle K
 \rangle}{K_{0}}$ is plotted against $\tilde \alpha$
for $N=8,16,32$ with the single fuzzy sphere start.
The horizontal line corresponds to the exact result.
}
  \label{sdeqFS}
}

The exact result (\ref{sde-quantity})
can be derived from the invariance of the partition
function (\ref{bosonicZ}) 
under the change of integration variables 
$A_\mu \mapsto (1 + \epsilon) A_\mu$.
Alternatively one may derive it from
\begin{eqnarray}
0 = \int \dd A \frac{\partial}{\partial A^{a}_{\mu}} 
\left\{ \tr(t^{a}A_{\mu}) \, \ee^{-S} \right\} \ ,
\label{SDE}
\end{eqnarray}
which is one of the Schwinger-Dyson equations
in this model.

In Fig.\ \ref{sdeqFS} we show our results for
the ratio $\frac{\langle K  \rangle}{K_{0}}$.
All the data
for various $\tilde{\alpha}$ and $N$
are consistent with `1' within error bars,
which demonstrates the validity of our simulation.

The exact result (\ref{sde-quantity})
is consistent with our assertion that
higher loop corrections are suppressed at large $N$
in the fuzzy sphere phase.
Using (\ref{one-loopf-sq}) and (\ref{one-loopcs-a}), we obtain
\beqa
 \langle K \rangle _{\rm 1-loop} 
 &=& \left\langle \frac{1}{N} \tr F^2 \right\rangle_{\rm 1-loop} 
  + 3 \, \alpha \, \langle M \rangle_{\rm 1-loop}   \non
 &=&  \frac{\tilde{\alpha}^{4}}{2} +  3 \, \tilde{\alpha}
\left( - \frac{{\tilde \alpha}^{3}}{6} + \frac{1}{{\tilde \alpha}} 
\right) = 3 \ ,
\label{oneloop-sde}
\eeqa
which agrees with the exact result (\ref{sde-quantity})
in the large $N$ limit.
We also notice that as $\tilde{\alpha}$ becomes large,
there is a huge cancellation between the first term 
$\langle \frac{1}{N} \tr (F_{\mu \nu})^{2} \rangle$ and 
the second term $3 \alpha \langle M \rangle$.
Namely the large $\tilde \alpha$ behavior of each term is
O($\tilde\alpha ^4$), but the leading term cancels exactly
leaving behind an O(1) quantity.
This is reflected in Fig.\ \ref{sdeqFS},
where the error bars become much larger
at $\tilde{\alpha} > \tilde{\alpha}_{\rm cr}^{\rm (l)}$.
Such a cancellation is absent 
in the Yang-Mills phase
since the second term $3 \alpha \langle M \rangle$ is close to zero
as one can see from Fig.\ \ref{miscFS}.

\section{The eigenvalue distribution of the Casimir operator}
\label{section:width}

In the previous Section we have seen that our Monte Carlo results
in the fuzzy sphere phase agree very well with the one-loop calculations 
around the single fuzzy sphere.
Here we show more directly that
the dominant configurations in this phase indeed 
have the geometry of the 2-sphere.

For that purpose it is convenient to consider the distribution
\beq
f (x) = \frac{1}{N} \sum_{j=1}^N
\left\langle \, \delta ( x - \lambda _j  ) \, \right\rangle \ ,
\label{distr_casimir}
\eeq
where $\lambda_j$ ($j=1,\cdots , N$) are $N$ real positive
eigenvalues of the Casimir operator
\beq
Q = \sum_{\mu} (A_\mu)^2 \ .
\label{casimir_op}
\eeq
Note that the set of eigenvalues $\{ \lambda_i \} $
is invariant under both SO(3) and SU($N$) transformations.
Within the classical approximation, the distribution $f(x)$ is given by
\beq
f(x) =  \delta  (x - R^2 ) 
\label{f_single}
\eeq
for the single fuzzy sphere (\ref{config_single}),
where $R$ is defined in (\ref{rad_single}).
Similarly for the multi fuzzy spheres (\ref{general_FS}) we get
\beq
f(x) =  \sum_{a=1}^k \, \left( \frac{n_a}{N} \right) \, \delta (x - r_a^2 ) 
\ ,
\eeq
where $r_a$ is defined in (\ref{rad_multi}).
This will play an important role in
identifying various multi fuzzy spheres in Section \ref{section:multi}.

    \FIGURE{\epsfig{file=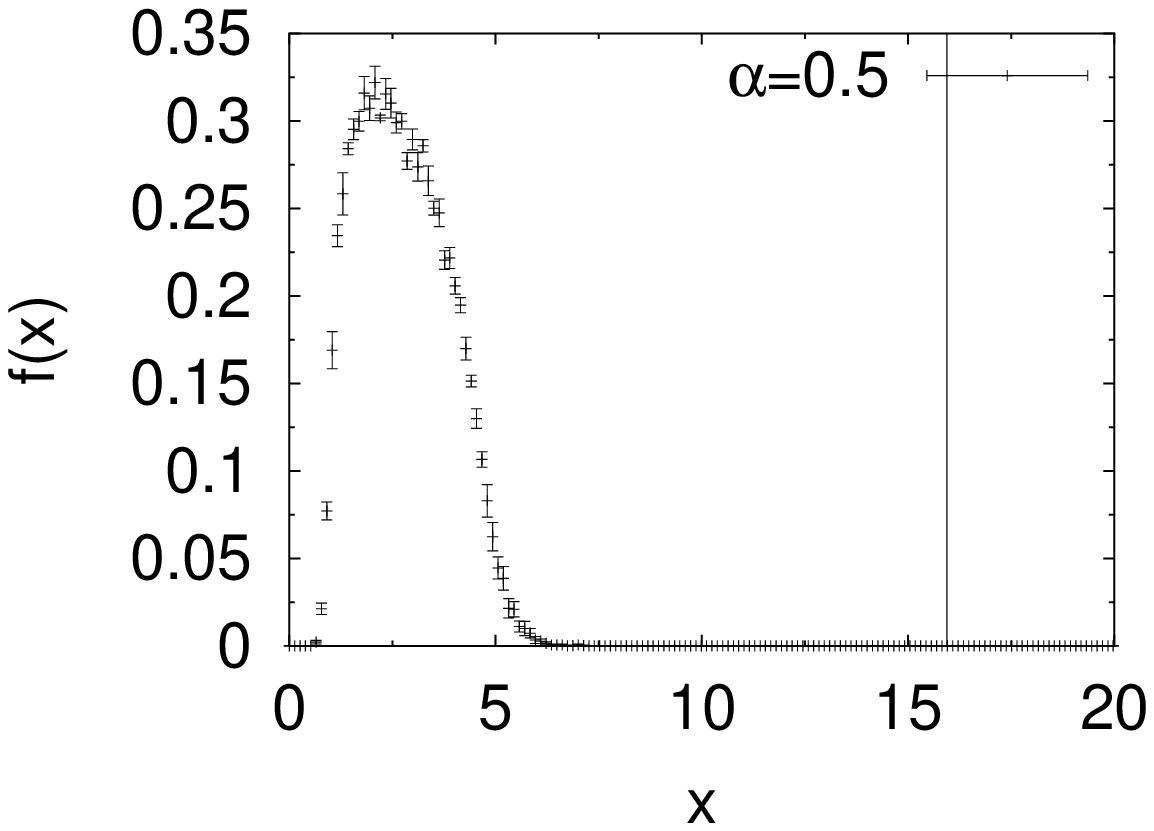,width=7.4cm}
            \epsfig{file=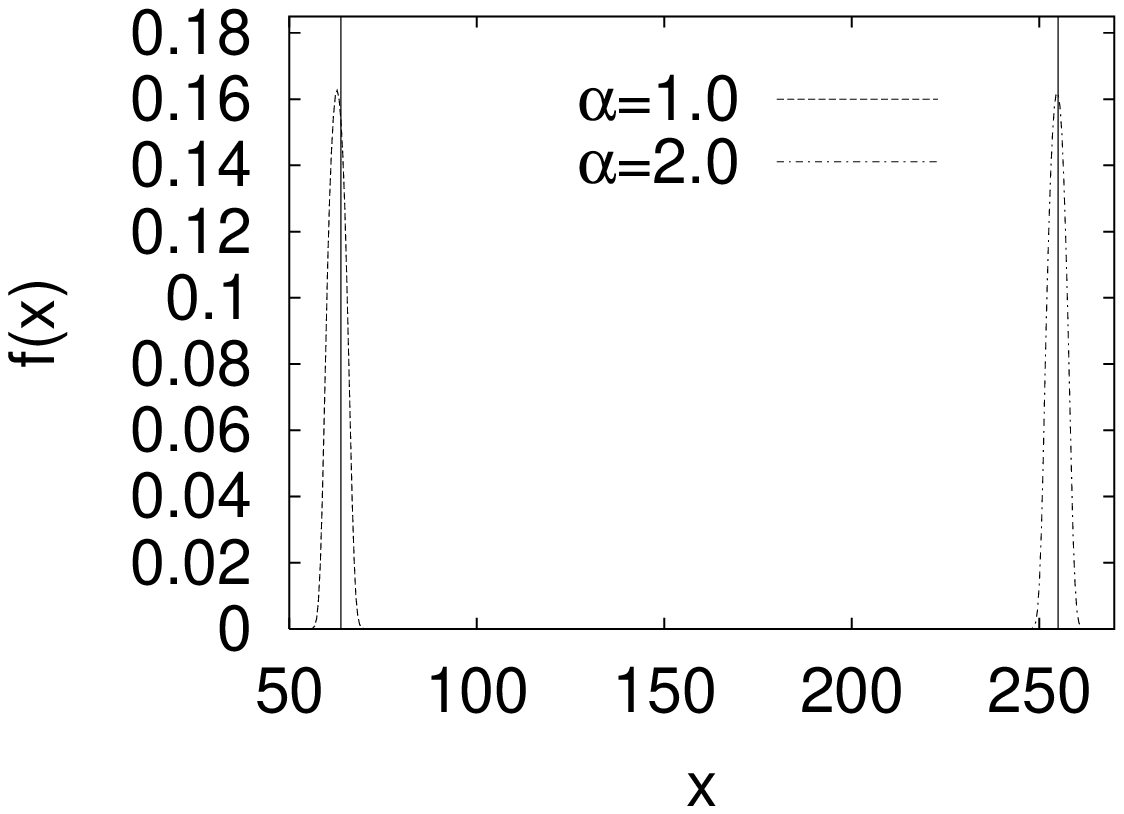,width=7.4cm}
    \caption{The eigenvalue distribution $f(x)$
of the Casimir operator (\ref{casimir_op})
is plotted for $N=16$ at
$\alpha=0.5 < \alpha_{\rm cr}^{\rm (l)}$ (left) 
and $\alpha=1.0, 2.0 > \alpha_{\rm cr}^{\rm (u)}$ (right).}
  \label{n16stability}}

Fig.\ \ref{n16stability} shows the results
for $\alpha=0.5 < \alpha_{\rm cr}^{\rm (l)}$ (left) 
and $\alpha = 1.0, 2.0 > \alpha_{\rm cr}^{\rm (u)}$ 
(right) at $N=16$.
The vertical lines represent the delta function
corresponding to the classical result (\ref{f_single}) for the 
single fuzzy sphere.
At $\alpha = 1.0$ and $2.0$ the measured distribution is close to the
classical result.
Slight deviations such as the smearing and the
shift can be understood as quantum effects, 
as we will see in Section \ref{stab}.
Thus the dominant configurations are not given {\em exactly} by
the single fuzzy sphere (\ref{config_single}), 
but they still preserve the geometrical structure
of a 2-sphere with a slightly smaller radius and with a finite width.

The result for $\alpha = 0.5$, on the other hand, deviates 
drastically from the classical result (\ref{f_single}).
In Section \ref{Yang-Mills} we will discuss
how the observed distribution 
can be understood qualitatively by
considering the diagonal configuration
(\ref{diagsolution}) and the effective action (\ref{eff_diag}).

At the intermediate values of $\alpha$
we observe a two-state signal, and as a result
the measured distribution is given by the superposition of the 
two types of distribution described above.
This is expected from the first-order nature of the phase transition
revealed in Section \ref{section:YangMills}.

\FIGURE{\epsfig{file=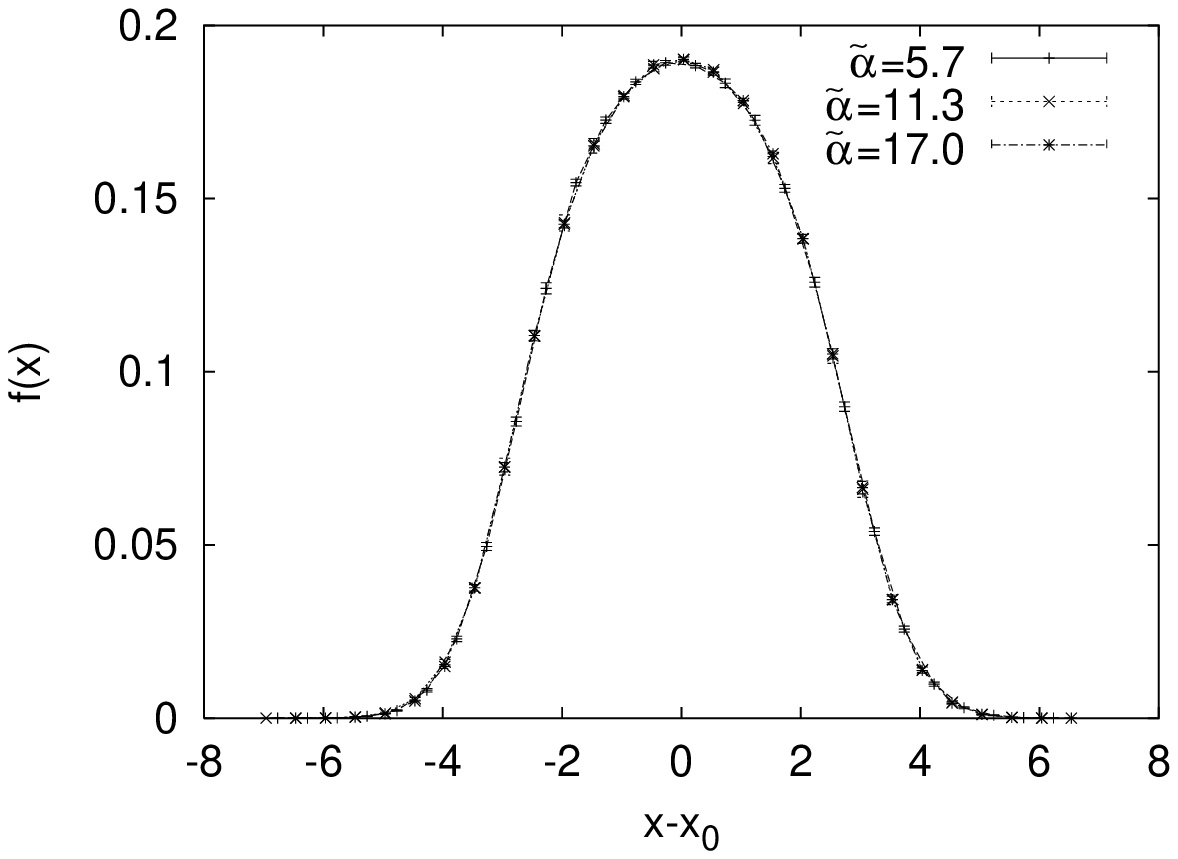,width=7cm}
        \epsfig{file=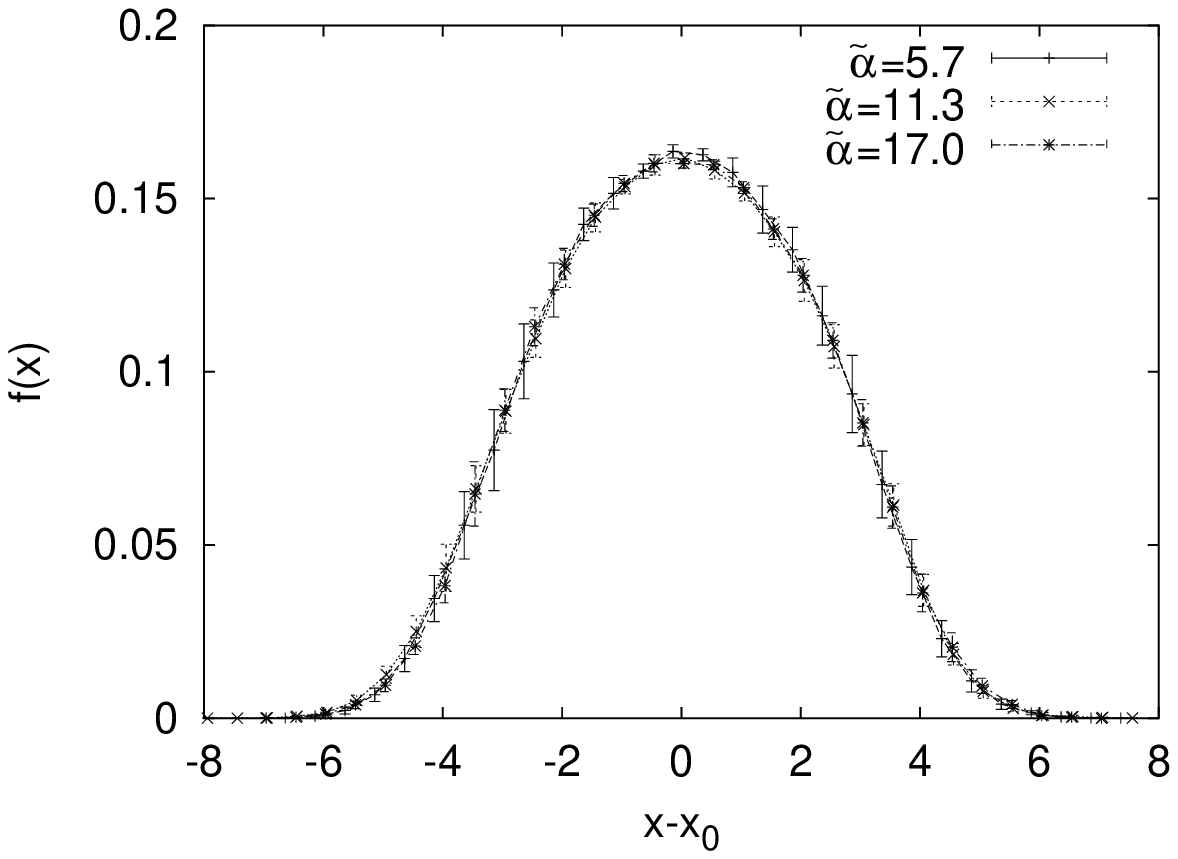,width=7cm} 
        \epsfig{file=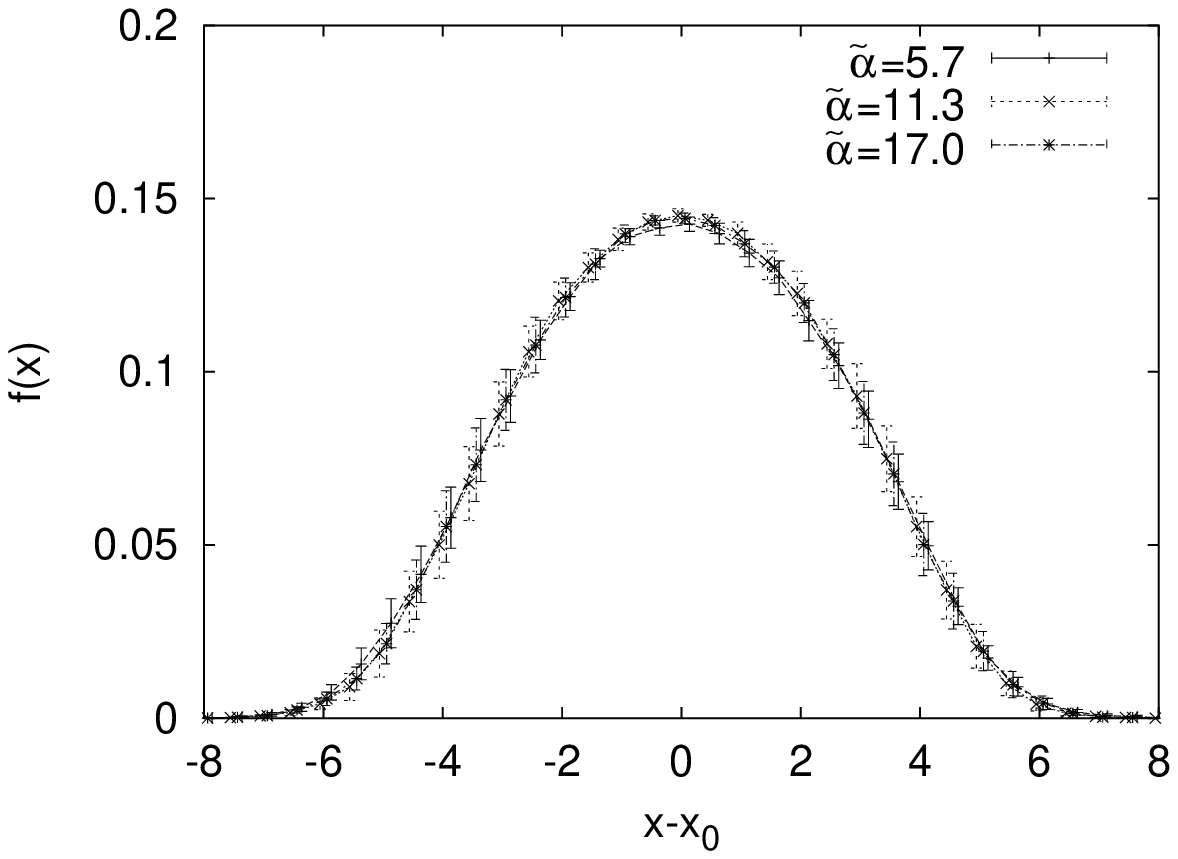,width=7cm}
        \epsfig{file=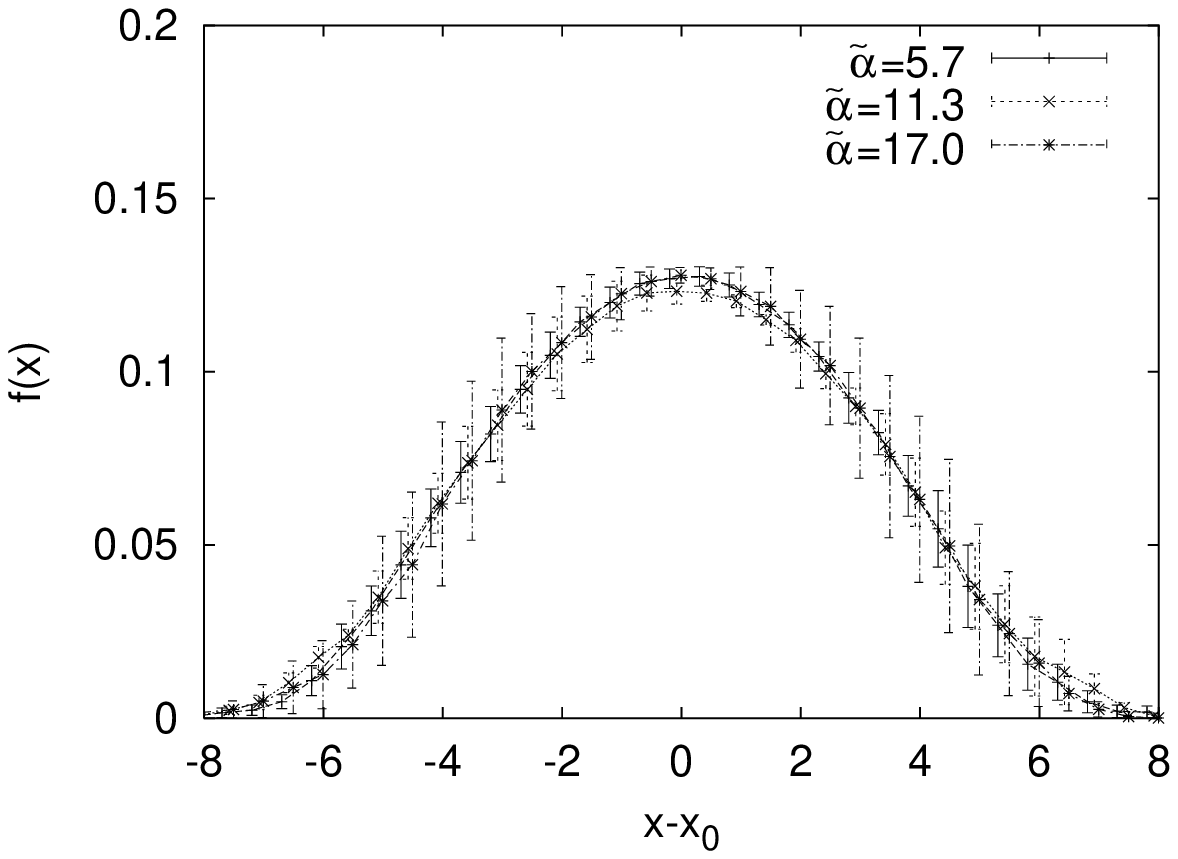,width=7cm} 
    \caption{The eigenvalue distribution $f(x)$ of the Casimir operator
(\ref{casimir_op}) is plotted against $x-x_0$ for $N=8$ (upper left), 
$N=16$ (upper right), $N=32$ (lower left) and $N=64$ (lower right),
respectively. 
At each $N$ the results for different $\tilde{\alpha}$ lie on
top of each other.
}
  \label{hists}}

\subsection{The distribution in the fuzzy sphere phase}
\label{stab}

Here we discuss the properties of the distribution $f(x)$
in the fuzzy sphere phase in more detail,
and show that they can be understood by the one-loop dominance.

\FIGURE{\epsfig{file=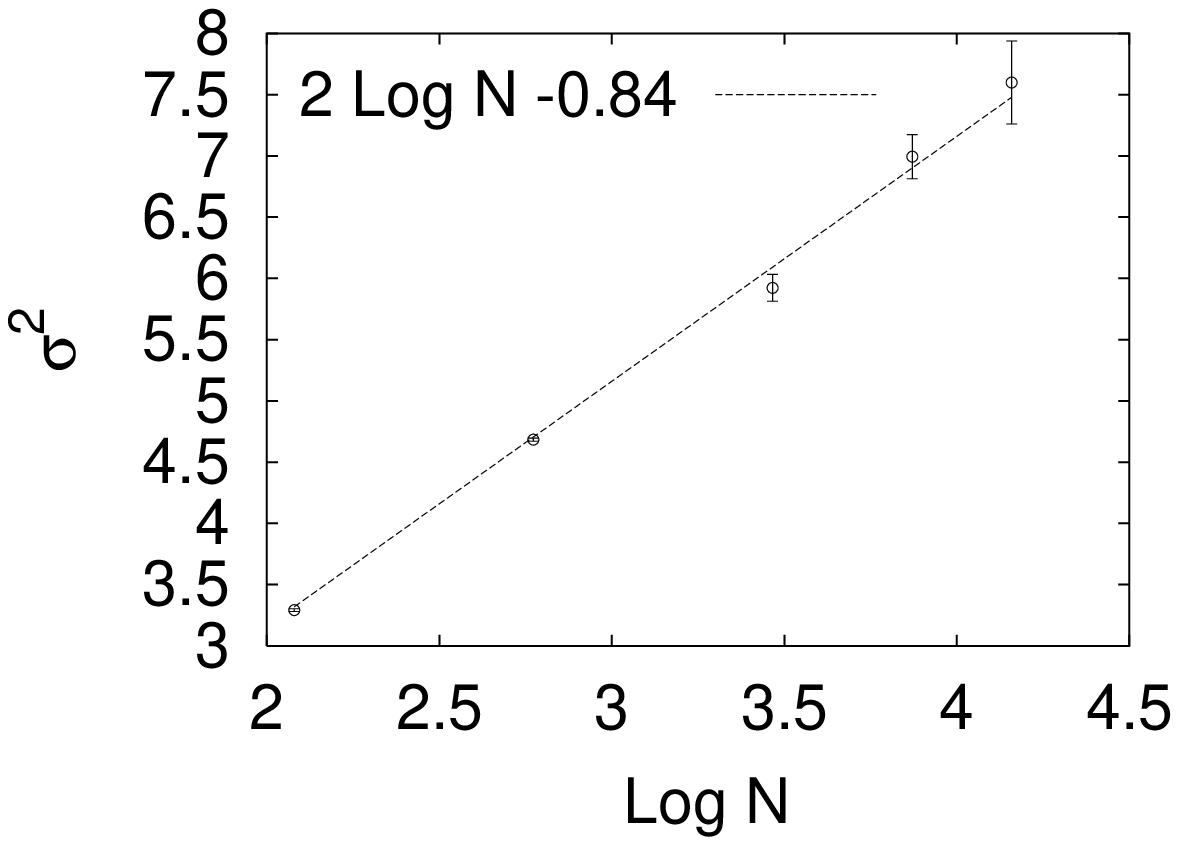,width=7.4cm}
    \caption{The variance $\sigma^{2}$ of the distribution
$f(x)$ is plotted for $N=8,16,32,48,64$.
The straight line is a fit to $\sigma^{2} \sim 2 \log N + b$,
where $b = - 0.84$.}
  \label{plot_width}}

Let us start with the position of the peak, which may be defined by
\beq
x_0 \equiv \int^{\infty}_{0} \dd x \,  x f(x)  
 = \left\langle \frac{1}{N} \tr (A_{\mu})^{2} \right \rangle  \ .
\label{peak}
\eeq
This quantity is actually studied in Section \ref{section:low_critical}
and it agrees with the one-loop result
(\ref{one-loopa-sq}) for the fuzzy sphere.
Note that the one-loop correction is negative, which 
implies that the peak moves toward the origin 
compared with the classical result (\ref{f_single}).
Also the one-loop correction 
is proportional to $\frac{1}{\tilde \alpha ^2}$,
meaning that the shift of the peak
becomes more pronounced at smaller $\tilde \alpha$.
These properties are indeed observed 
in Fig.\ \ref{n16stability} (right).
An interesting point to note here is that 
the one-loop result (\ref{one-loopa-sq})
for $\frac{1}{N}\langle \frac{1}{N} \tr (A_{\mu}^{2}) \rangle$
becomes zero at $\tilde \alpha = \sqrt{2}$.
Assuming the one-loop dominance, we may suspect that
nonperturbative effects should become non-negligible 
as we decrease $\tilde \alpha$, which invalidates 
the perturbation theory before $\tilde \alpha$ reaches $\sqrt{2}$.
This is indeed what happens in reality
(Note that $ \tilde \alpha _{\rm cr}^{\rm (l)} = 2.1 > \sqrt{2}$).

Having understood the position of the peak,
let us turn to the shape of the distribution.
In Fig.\ \ref{hists} we
plot the distribution $f(x)$ against $x-x_{\rm 0}$
for various $\tilde \alpha$ at $N=8,16,32,64$.
The results for different $\tilde \alpha$
collapse to a single curve, which means that
the shape of the distribution does not depend on $\tilde \alpha$.
Furthermore the distribution turns out to be symmetric around $x=x_0$.
These properties can be understood from the `one-loop dominance'.

Among the quantities that characterize the shape of distributions,
let us focus on the width $\sigma$, which may be defined by
\beqa
   \sigma^{2} &\equiv& \int^{\infty}_{0} \dd x \, (x - x_0)^{2} f(x) 
\non
    &=& \left\langle \frac{1}{N} \tr ({A_{\mu}}^{2})^{2} \right\rangle
    - \left\langle \frac{1}{N} \tr (A_{\mu})^{2} \right \rangle^2
\ .
    \label{sigma2_def}
\eeqa
We can calculate it by the perturbative expansion around 
the single fuzzy sphere.
Since the classical result is zero, it starts from
the one-loop contribution, which is given by
\beq
\sigma ^2 = 2 \log N  + {\rm O}(1)
\label{sigma-1-loop}
\eeq
at large $N$ (See Appendix \ref{one-loop-obs-appendix} for
a derivation).
The perturbative result is independent of $\alpha$ at this order,
but higher order corrections may yield terms proportional to
$(\frac{1}{\alpha^4})^{n}$ ($n=1,2,\cdots$).
The $\tilde \alpha$ independence observed in our Monte Carlo data
can therefore be understood by the one-loop dominance
at large $N$.
In Fig.\ \ref{plot_width} we plot the variance (\ref{sigma2_def})
obtained from simulations with $N = 8,16,32,48,64$.
Taking account of the O(1) term, we find that
the large $N$ behavior
agrees very well with the one-loop result (\ref{sigma-1-loop}).


\subsection{Space-time picture in the Yang-Mills phase}
\label{Yang-Mills} 

Let us move on to the result for the eigenvalue distribution $f(x)$
in the Yang-Mills phase, which is shown
in Fig.\ \ref{n16stability} (left).
If the dominant configurations in this phase 
were given by the diagonal configurations (\ref{diagsolution}),
the eigenvalues of the Casimir operator
would be given by $\sum_{\mu} (x_\mu^{(j)})^2$ ($j=1,\cdots , N$).
Therefore the eigenvalue distribution $f(x)$ of the Casimir operator 
(\ref{casimir_op})
would be related to the radial density distribution $\rho(r)$ 
of the $N$ space-time points $x_\mu^{(j)}$ ($j=1,\cdots , N$) by
\beq
f(x) \, \dd x = \rho (r) \, 4 \pi r^2 \, \dd r \ ,
\label{f_to_rho}
\eeq
where $x=r^2$.
From (\ref{f_to_rho}) we get
\beq
\rho(r) = \frac{1}{2 \pi r } f(r^2 )  \ .
\label{def_rho}
\eeq
In reality the dominant configurations are {\em not} diagonal
up to the SU($N$) symmetry,
but we may still define the `radial density distribution' $\rho(r)$
through (\ref{def_rho}).
In Fig.\ \ref{hist-anYM} we plot $\rho(r)$
for $N=8,16,32$ at $\alpha=0.0$ and for $N=32$
at $\alpha = 0.0 , \cdots , 0.6$. 
For $\alpha \ge 0.4$
we have to use the zero start in order to stay in the Yang-Mills phase.
The distribution at $\alpha = 0$ converges at large $N$,
and this is confirmed also for $\alpha \neq 0$.

\FIGURE{\epsfig{file=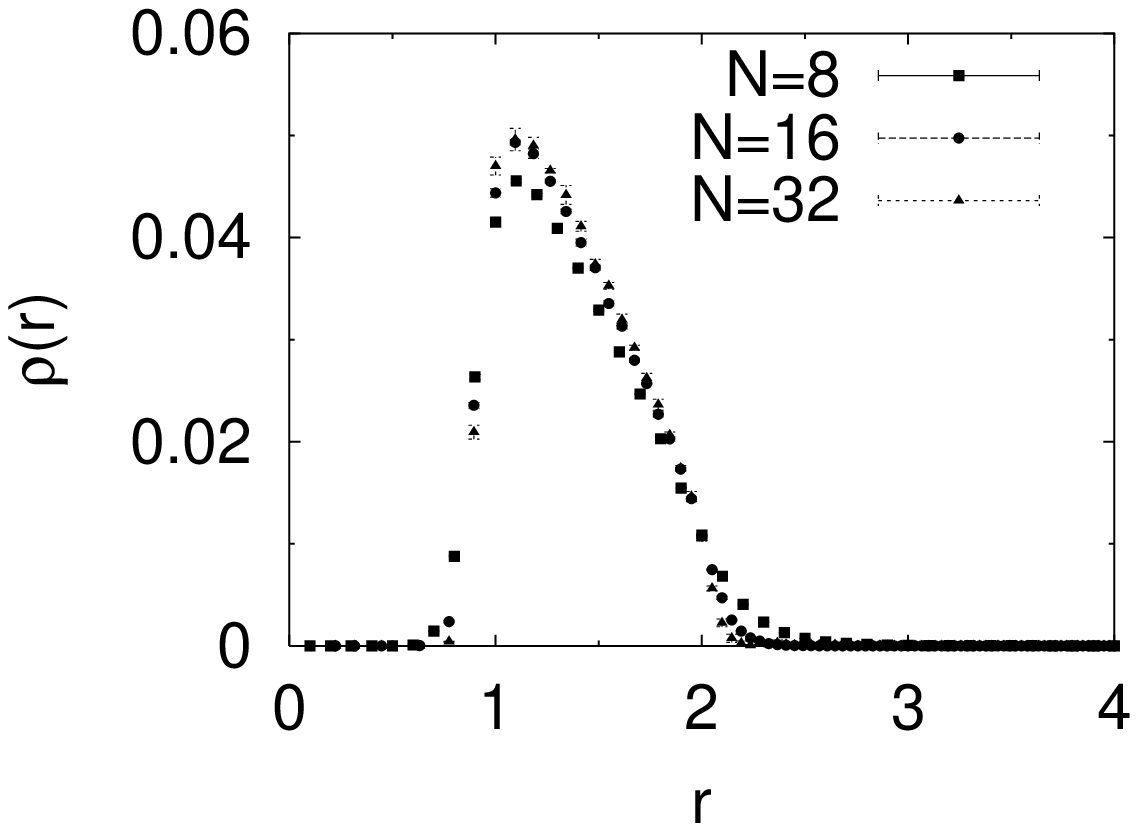,width=7.4cm}
           \epsfig{file=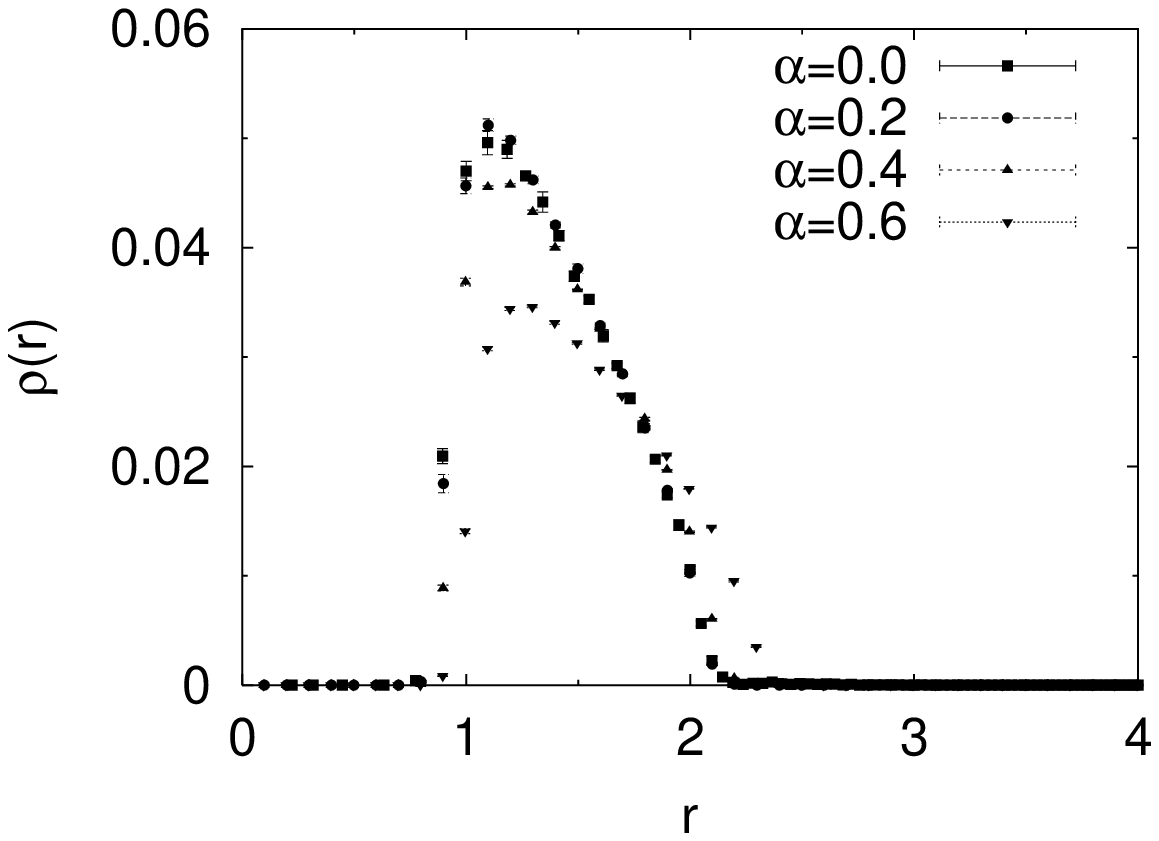,width=7.4cm} 
\caption{The `radial density distribution' $\rho(r)$
is plotted for $N=8,16,32$ at $\alpha=0.0$ (left)
and for $N=32$ at $\alpha = 0.0, 0.2, 0.4, 0.6$ (right).} 
\label{hist-anYM}}

An interesting point to note here is that
the distribution $\rho (r)$ has an empty region
$0 \le r \le r_0$, where $r_0 =0.8\sim 0.9$.
This may be understood as follows.
If there exists an eigenvector of the Casimir operator 
$Q = \sum_\mu (A_\mu)^2 $ whose eigenvalue is close to zero,
the same vector is an approximate eigenvector of 
$A_{1}$, $A_{2}$ and $A_{3}$ separately with eigenvalues
which are also close to zero.
Existence of such a vector is not allowed, however, since
a generic configuration $\{A_{\mu};\mu = 1,2,3\}$
that appears in the ensemble does not commute with each other. 
The appearance of the empty region can thus be understood 
by the `uncertainty principle'.
Note that the scale of noncommutativity is of O(1) since
$\langle \frac{1}{N} \tr F_{\mu \nu}^{2} \rangle \sim {\rm O}(1)$
in the Yang-Mills phase.
Therefore the size $r_0$ of the empty region should be of O(1) as well,
which is in agreement with our Monte Carlo results.

Except for this empty region, 
the distribution $\rho(r)$ decreases monotonously.
This behavior can be understood by 
the one-loop effective action (\ref{eff_diag}),
which gives rise to an attractive potential between
any pair of space-time points.
The rapid fall of $\rho(r)$ at large $r$ is consistent with
the results obtained for $\alpha = 0$ \cite{Krauth:1999qw}.
Note, however, that the precise form of the distribution cannot
be understood solely from the perturbation theory around the diagonal 
configuration (\ref{diagsolution}),
since the higher order corrections become non-negligible
as the space-time points come closer to each other due to the
one-loop potential (\ref{eff_diag}).
For instance the distribution $\rho(r)$ 
is seen to expand to larger $r$ as $\alpha$
increases, but this behavior does not agree with the naive
expectation from the one-loop effective action (\ref{eff_diag}),
where the second term adds to the attractive potential.
We may understand it from a nonperturbative point of view,
however, since the Chern-Simons term, which can be more negative 
for larger $A_\mu$,
effectively enhances configurations with a larger extent.

\section{Properties of the multi fuzzy spheres}
\label{section:multi}

As we described in Section \ref{section:def_model}
the model has various multi fuzzy spheres 
as classical solutions in addition to the single fuzzy sphere.
The aim of this Section is to see how they appear 
in simulations and to study their properties.

\subsection{Meta-stable states in the thermalization process} 
\label{evo}

   \FIGURE{\epsfig{file=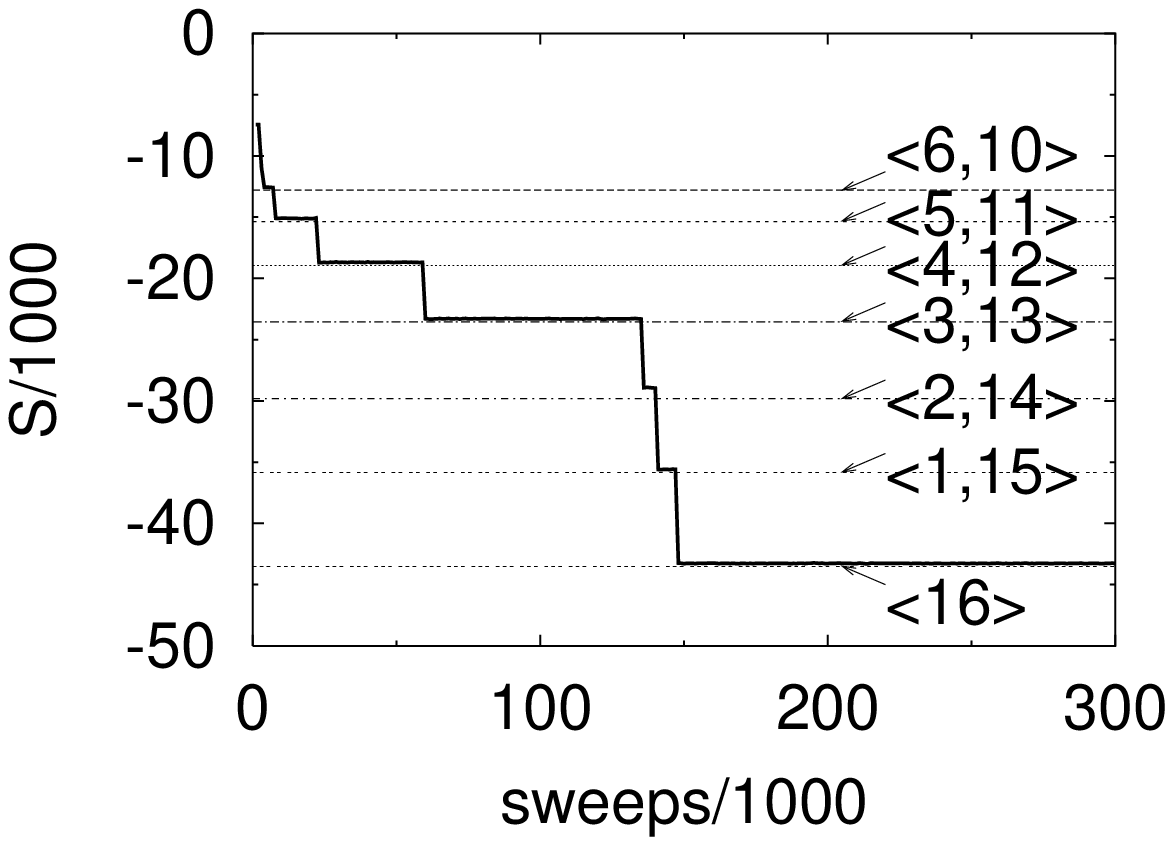,width=7cm}
           \epsfig{file=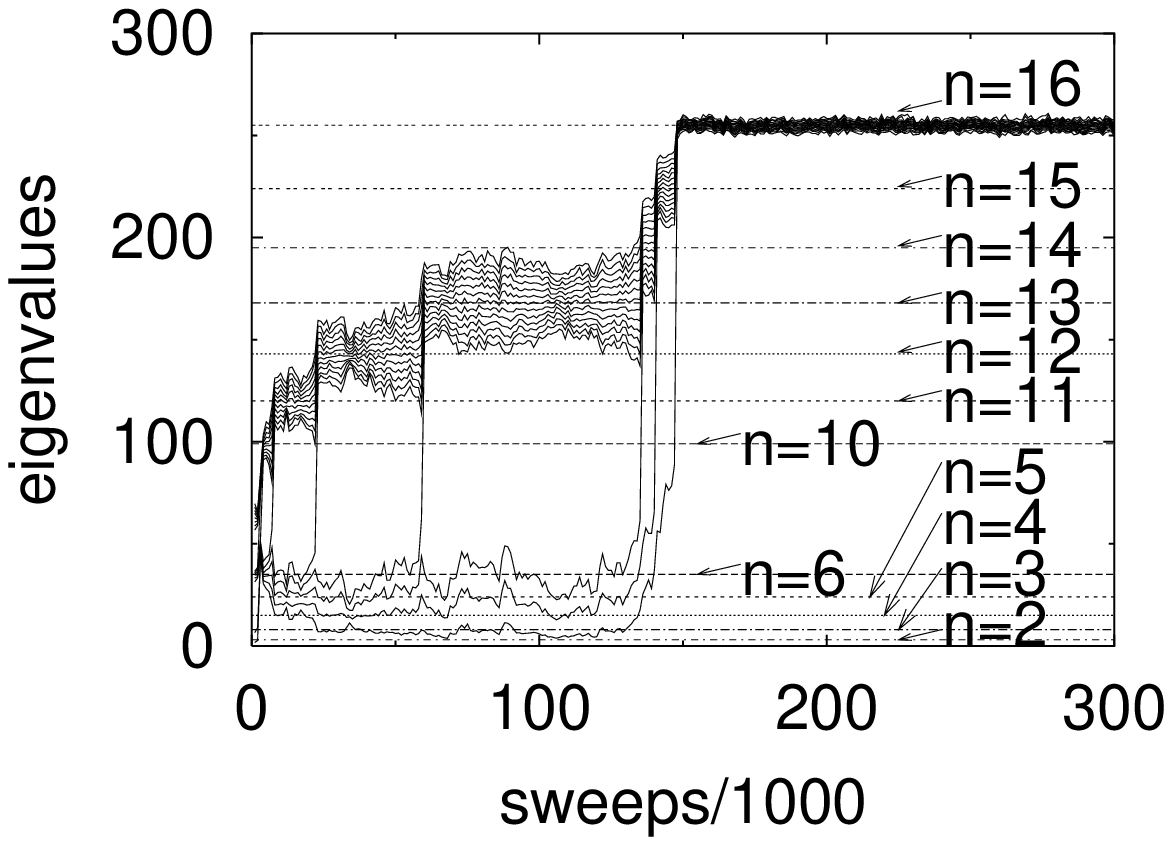,width=7cm}
    \caption{The history of the action 
  $S$ (left) and the eigenvalues of the Casimir operator (right)
  are plotted for $N=16$, $\alpha=2.0$ with the zero start.
The horizontal lines represent the classical results
for the action
and the radius squared, respectively.}
  \label{n16a2}}

Let us note first that the
action for the classical solution (\ref{general_FS})
is given by
   \begin{eqnarray}
    S = - \frac{\alpha^{4} N}{24} \sum_{a=1}^{k} (n_{a}^{3} -
    n_{a}) \ ,
   \label{multiaction}
   \end{eqnarray}
which becomes minimum for the single fuzzy sphere ($k=1$)
(Notice the constraint (\ref{summing_block})).
Therefore at the classical level the multi fuzzy spheres are
expected to appear only as meta-stable states.
\footnote{At the quantum level the issue becomes more nontrivial;
see Section \ref{comp_eff}.}

 \FIGURE{\epsfig{file=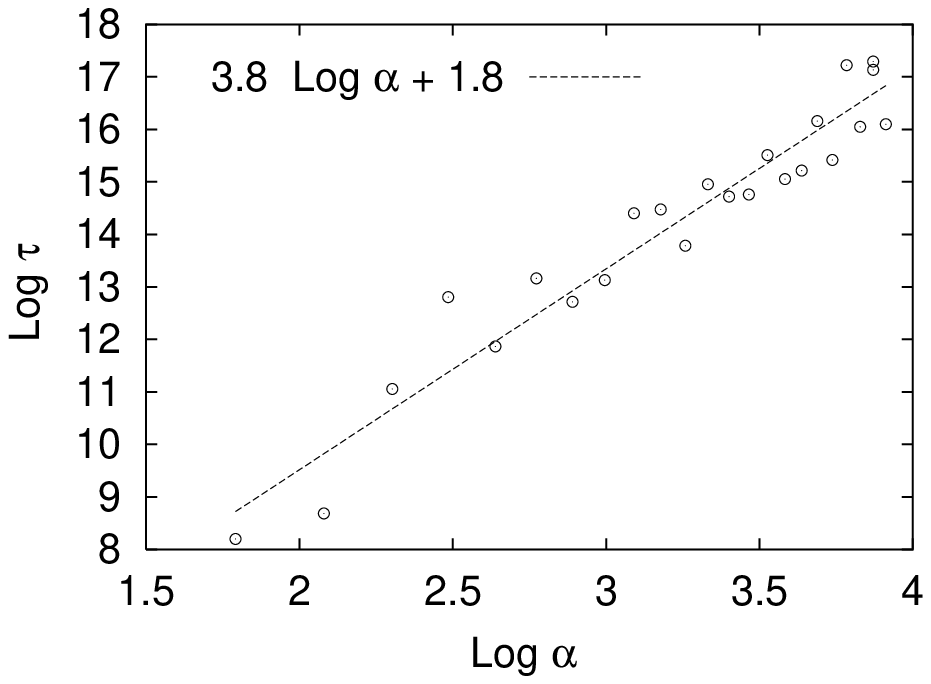,width=7.4cm}
    \caption{The `life time' ($\tau$) 
of the multi fuzzy spheres $ \langle 3,5 \rangle $ for $N=8$ 
is shown as a function of $\alpha$  in the log-log plot.
The straight line represents a fit to the power law.
}
  \label{life8}}

Indeed we do find them in the process of thermalization.
We have performed simulations at
$N=16$ and $\alpha=2.0 > \alpha _{\rm cr}^{\rm (u)}$
with the zero start.
Fig.\ \ref{n16a2} shows the history of the action $S$ and the 
eigenvalues of the Casimir operator (\ref{casimir_op}).
We have introduced a short-hand notation 
$ \langle  n_1$, $n_2$, $\cdots$, $n_k \rangle$
for the classical solution (\ref{general_FS}).
One sweep is defined as updates of all the elements of the 
matrices $A_\mu$ by the heat bath algorithm (See Appendix \ref{heat-bath}).
We see many plateaus before we reach the single fuzzy sphere
after 150,000 sweeps. These meta-stable states can be identified with
the multi fuzzy spheres by comparing the action
and the eigenvalues with the classical results 
(\ref{multiaction}) and (\ref{rad_multi}), respectively.

Next we start the simulation with a particular multi-fuzzy-sphere
configuration, and consider its `life time' defined
by the number of sweeps necessary for the state 
to decay into another state.
In practice the `life time' can be extracted
from the history of the action,
which has a clear plateau as the ones in Fig.\ \ref{n16a2} (left).
In Fig.\ \ref{life8} we show the `life time'
of the multi-fuzzy-sphere state $\langle 3,5 \rangle$ for $N=8$ 
as a function of $\alpha$ in the log-log plot.
Although the `life time' thus defined is a probabilistic quantity
depending on the particular series of random numbers,
we see from Fig.\ \ref{life8} that
it obeys a power law on the average.
The meaning of this observation shall be discussed in the next Section.

\subsection{The `life time' of the $k$ coincident fuzzy spheres} 
\label{life}

  \FIGURE{\epsfig{file=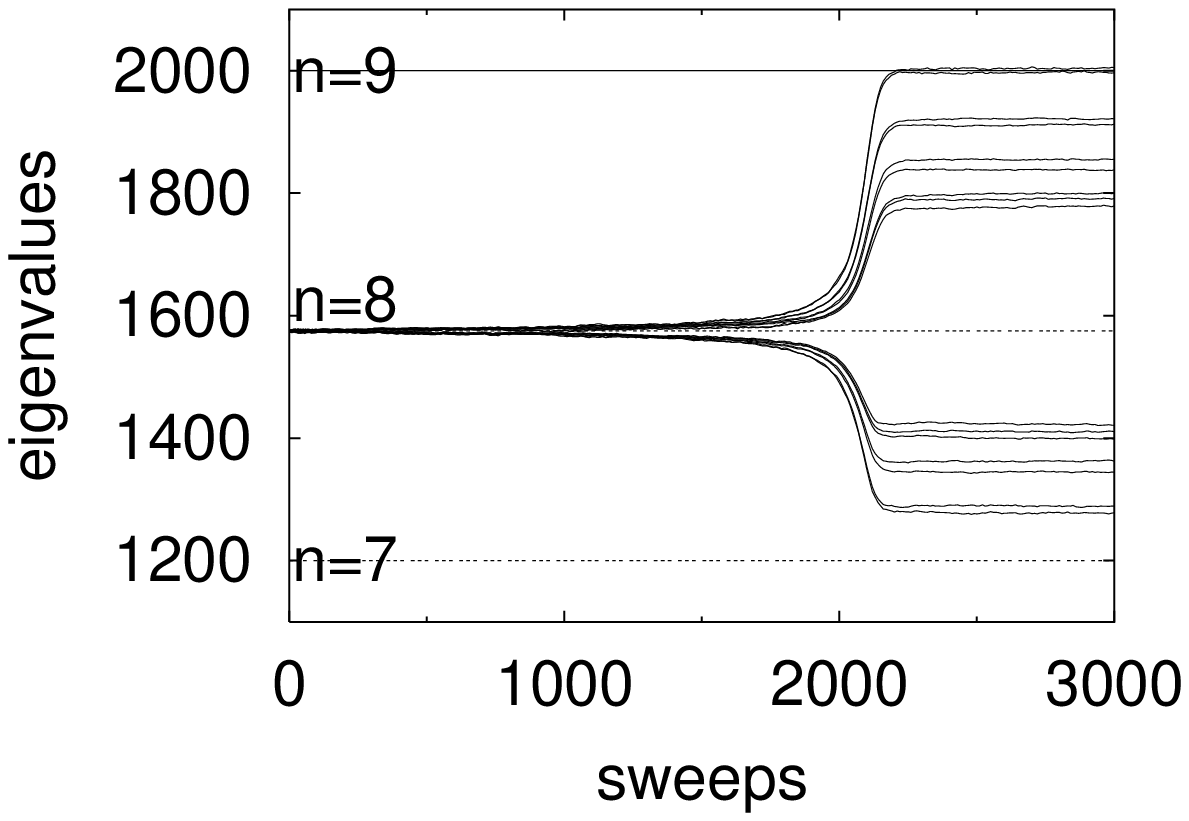,width=4.8cm}
          \epsfig{file=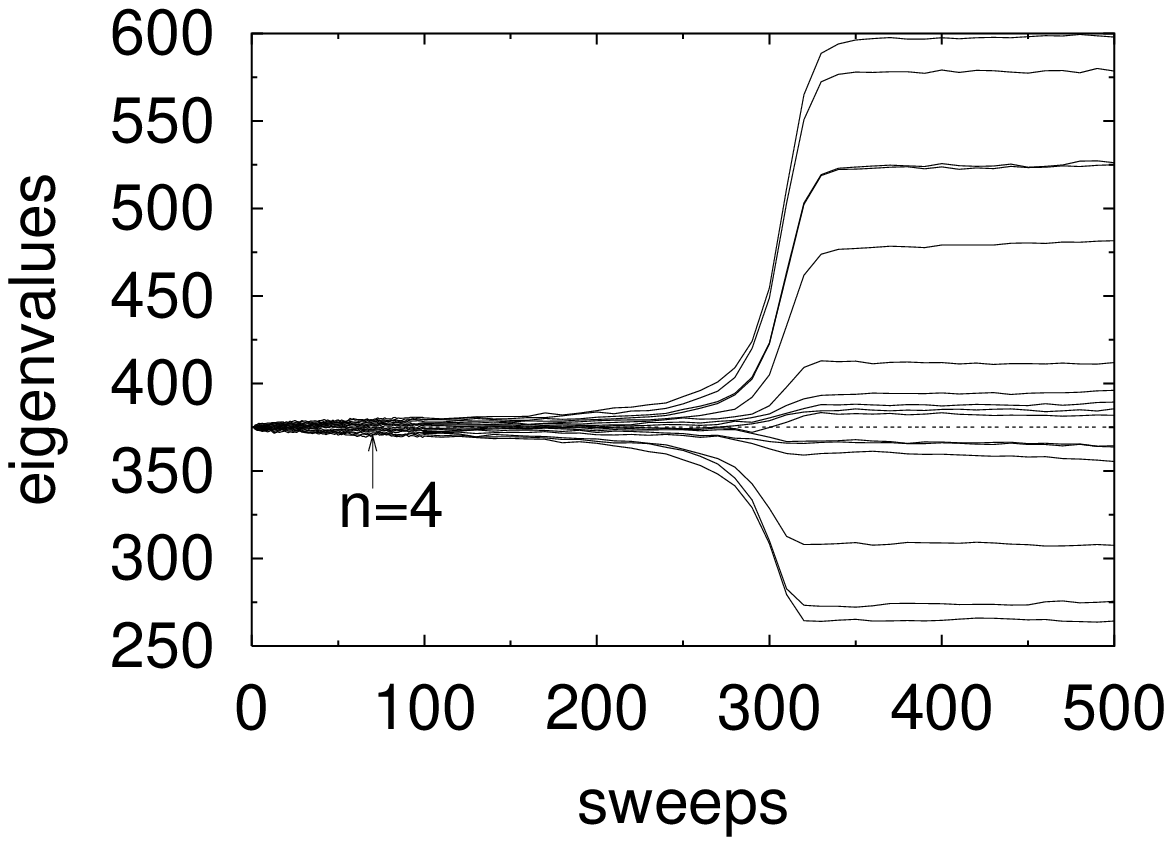,width=4.8cm}
          \epsfig{file=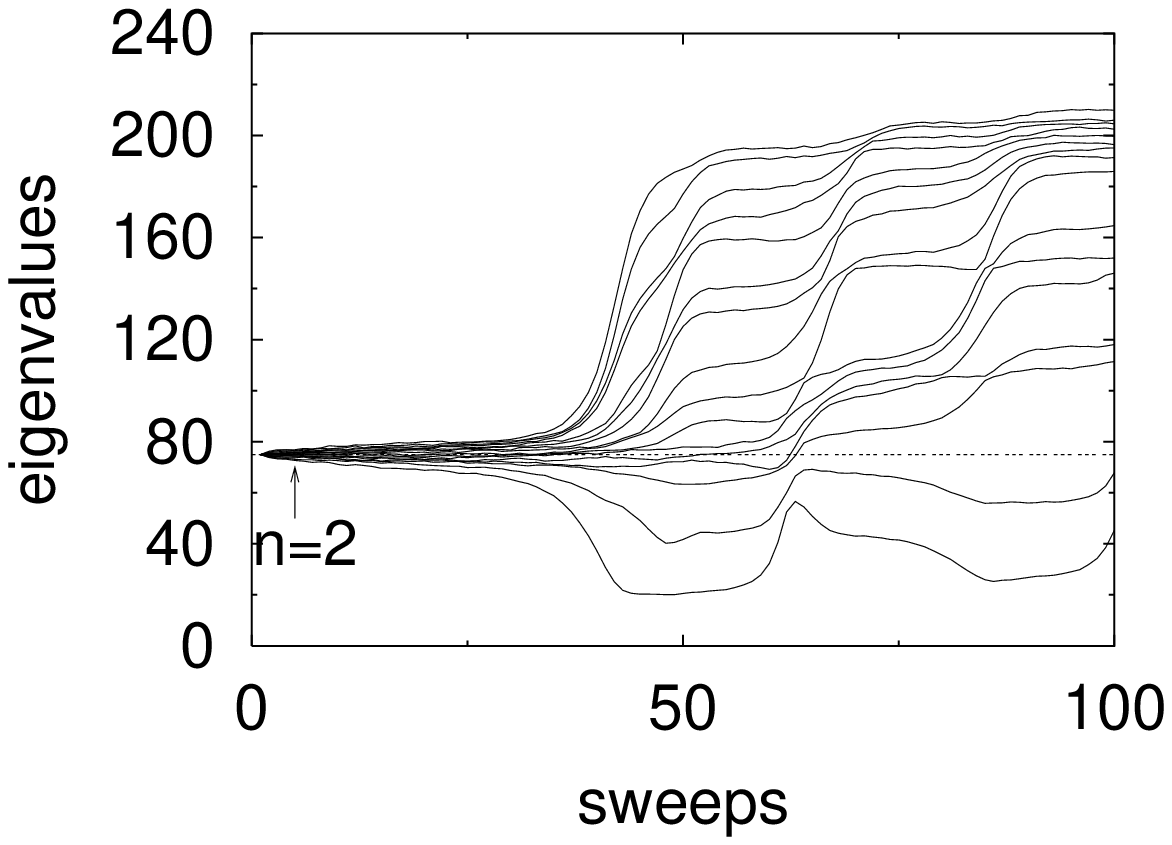,width=4.8cm}
 \caption{The history of the eigenvalues 
of the Casimir operator (\ref{casimir_op}) 
is plotted for $N=16$, $\alpha=10.0$.
The initial configuration is taken to be 
the $k$ coincident fuzzy spheres (\ref{kini}) 
with $k=2$ (left), $k=4$ (middle) and $k=8$ (right).} 
\label{decay248}}

From now on we discuss the properties of 
the $k$ coincident fuzzy spheres
\begin{eqnarray}
A_{\mu} = \alpha L_{\mu}^{(n)} \otimes {\bf 1}_{k} \ ,
\label{kini} 
\end{eqnarray}
which correspond to the case 
$n_1 = \cdots = n_k \equiv n \left( =  \frac{N}{k} \right)$
of (\ref{general_FS}).
Such a configuration 
is of particular interest since
it gives rise to a NC gauge theory on the fuzzy sphere 
with the gauge group of rank $k$.
From the viewpoint of perturbative calculations,
the $k$ coincident fuzzy spheres are easier to handle than
more general multi fuzzy spheres, and we have obtained explicit results
in the Appendices \ref{eff_derive} and \ref{one-loop-obs-appendix}.
Here we perform Monte Carlo simulation with 
$N=16$, $\alpha=10.0$ starting from 
the $k$ coincident fuzzy spheres (\ref{kini}) with $k=2,4,8$.
Fig.\ \ref{decay248} shows the history of the eigenvalues
of the Casimir operator.

   \FIGURE{\epsfig{file=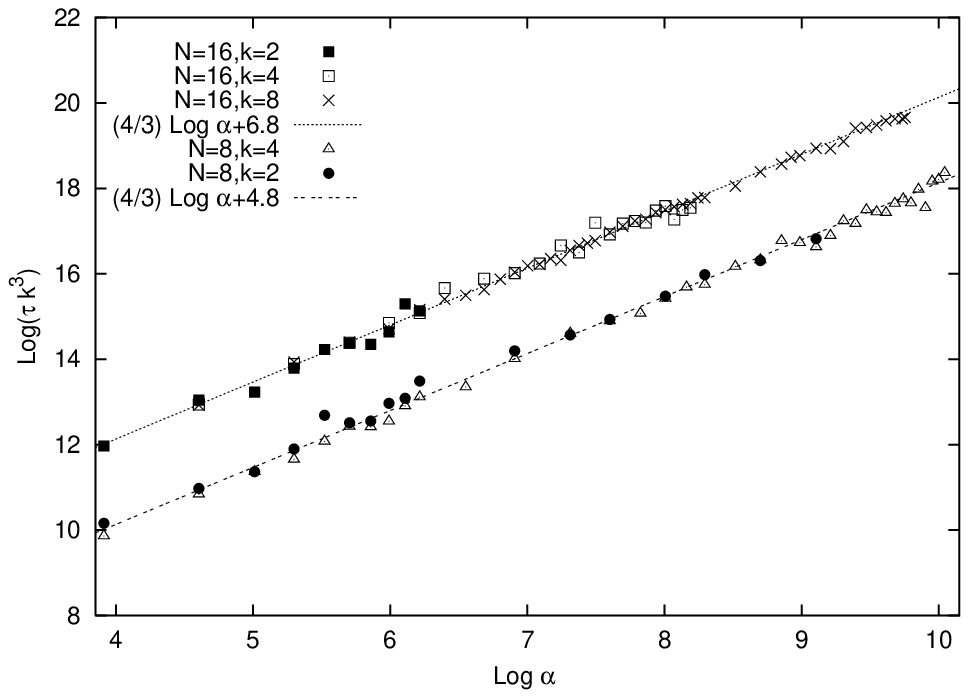,width=12cm}
    \caption{The `life time' ($\tau$) 
of $k$ coincident fuzzy spheres with $k=2,4,8$ follows
a universal power law. The plot shows
$\log( \tau \, k^{3} )$ v.s.\ $\log \alpha$ for $N=8,16$.
The straight lines represent the fits to
$\log( \tau  \,  k^{3} ) = \frac{4}{3} \log \alpha + c$.
}
  \label{life16}}

We observe that the $k$ coincident fuzzy spheres also
have a finite `life time', and it becomes shorter for larger $k$.
For all $k$ the decay process starts with the spreading of the eigenvalues.
We have continued the runs shown in Fig.\ \ref{decay248} until
the configurations finally thermalize. 
After visiting various multi fuzzy spheres on the way,
all the simulations starting with different $k$ end up with the
single fuzzy sphere, the final stage looking very much like
Fig.\ \ref{n16a2}.
The `life time' $\tau$
can be extracted from the plateau in the history of the action
as in the previous Section.
In Fig.\ \ref{life16} we plot
$\log( \tau \,  k^{3} )$ against $\log \alpha$,
which reveals a power law
\beq
\tau \propto \alpha^{\frac{4}{3}} \,  k^{-3}
\label{power_law}
\eeq
for both $N=8$ and $N=16$ over a huge range of $\alpha$
($\alpha = 50 \sim 20000$).
For a different algorithm, the `life time' will be 
multiplied by some constant which is naively expected to be
independent of both $\alpha$ and $k$.
If that is the case, the powers obtained here should have some
universal meaning.

This result is in striking contrast to the naive expectation that the
decay probability $P$ is given by $P \sim \ee^{-S} = \exp \left(
- {\rm const.} \frac{N^{4} \alpha^{4}}{k^{2}} \right)$ and
the `life time' is given by its inverse.
We interpret it as a consequence of
certain instability 
as discussed in the Appendix \ref{instability}.

The power law observed for a more general 
multi-fuzzy-sphere state $\langle 3,5 \rangle$
in the previous Section
may be understood as a consequence of
the instability for shifting the center of the spheres
relatively.
This is consistent with
the rather smeared distribution of the eigenvalues 
observed in Fig.\ \ref{n16a2} (right) as compared with
the distribution for the single fuzzy sphere.
Such instability, however, may disappear at larger $N$
as discussed for the coincident case in the Appendix \ref{instability}.
Then the decay of multi fuzzy spheres will be suppressed exponentially.
Indeed at $N=32$ we observed some cases where a multi-fuzzy-sphere
state does not decay into other states during the simulations.

\subsection{The one-loop dominance for the $k$ coincident fuzzy spheres} 

Although the $k$ coincident fuzzy spheres 
have a finite `life time',
we can still measure the observables studied in
Section \ref{section:low_critical}
and calculate the 
expectation value before the decay actually occurs.
The observables can also be calculated within the perturbation theory
around the $k$ coincident fuzzy spheres (\ref{kini})
neglecting the zero modes (See Appendix C).
The leading large $N$ results at the one-loop level is given by 
(\ref{tr_A^2_largeN}), (\ref{o2_largeN}), (\ref{o1}) and 
(\ref{appendix_(S)}).


  \FIGURE{\epsfig{file=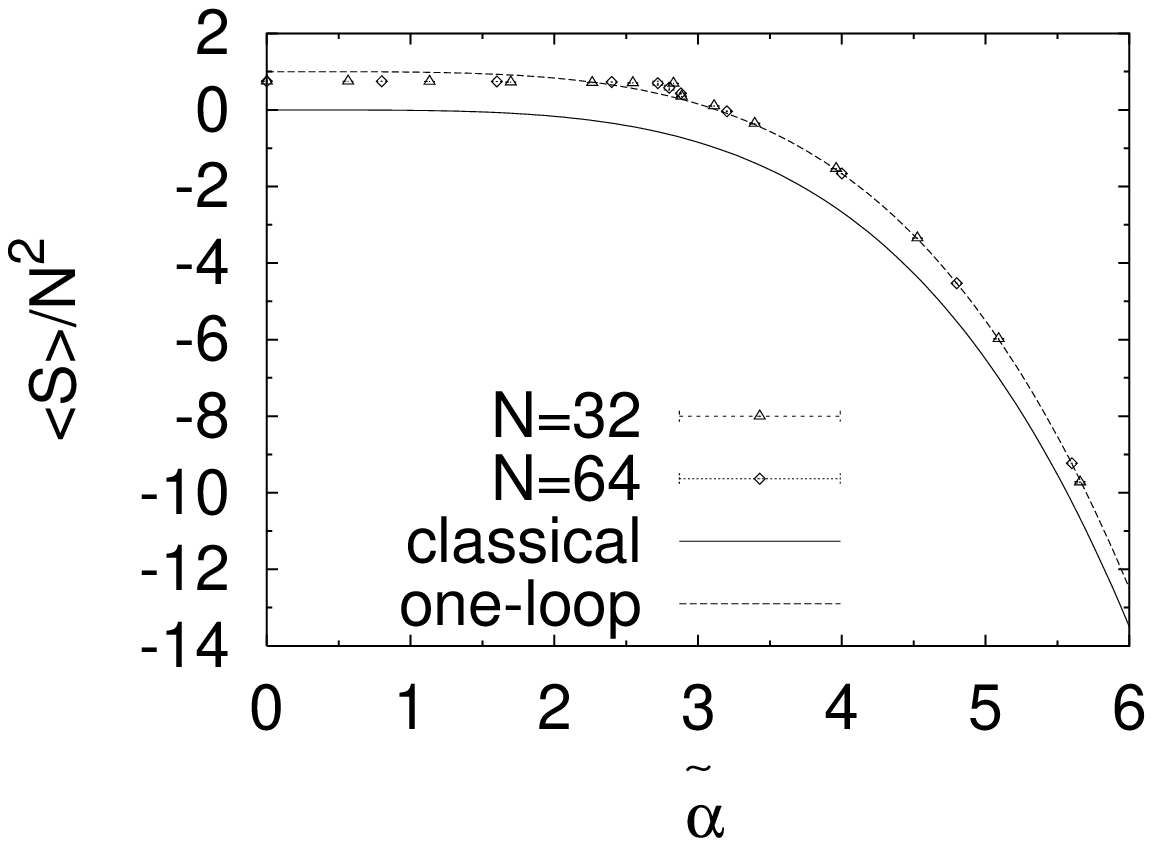,width=7.4cm}
          \epsfig{file=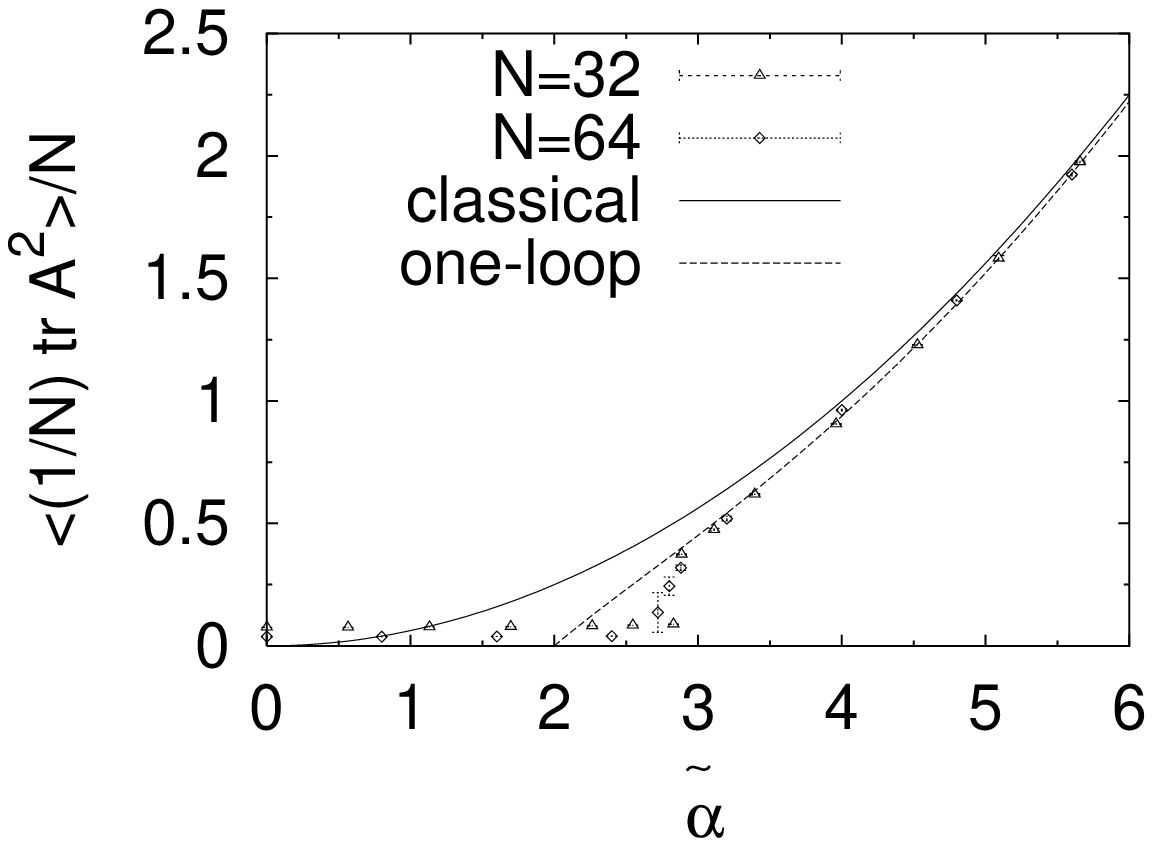,width=7.4cm}
          \epsfig{file=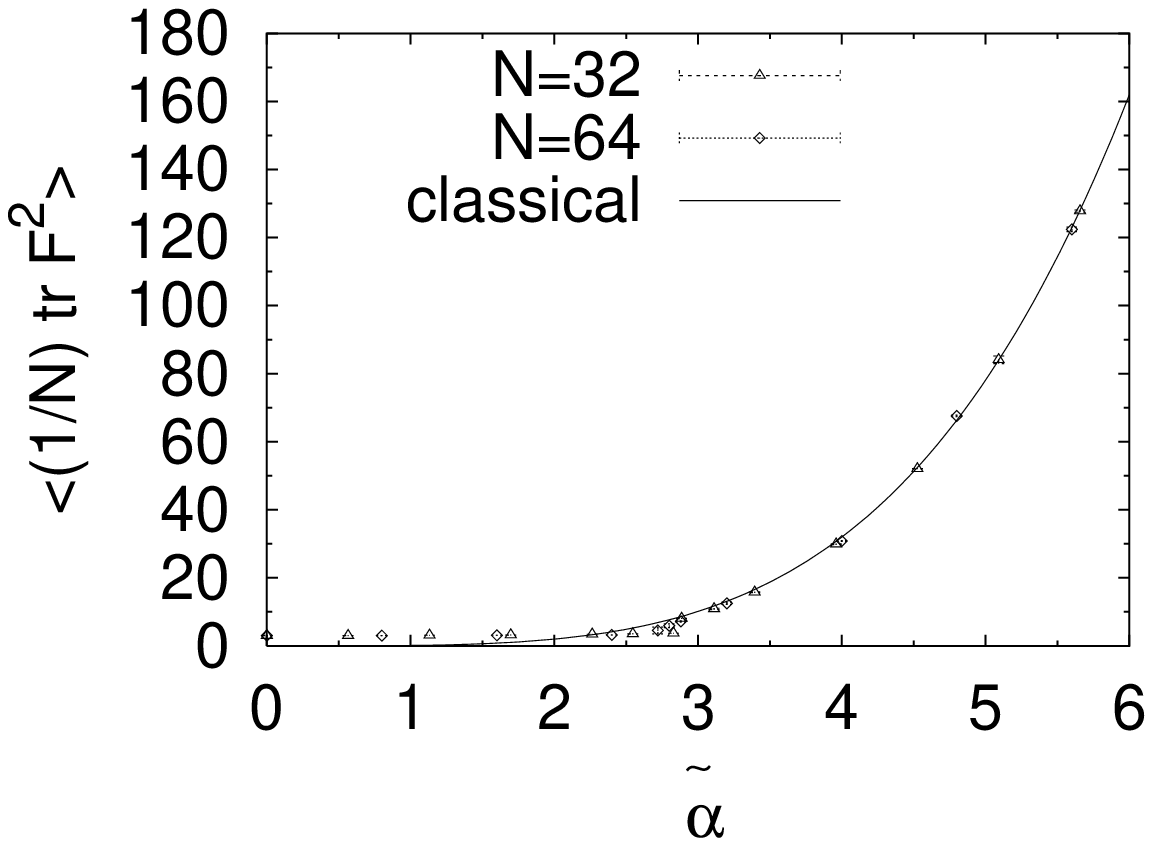,width=7.4cm}
          \epsfig{file=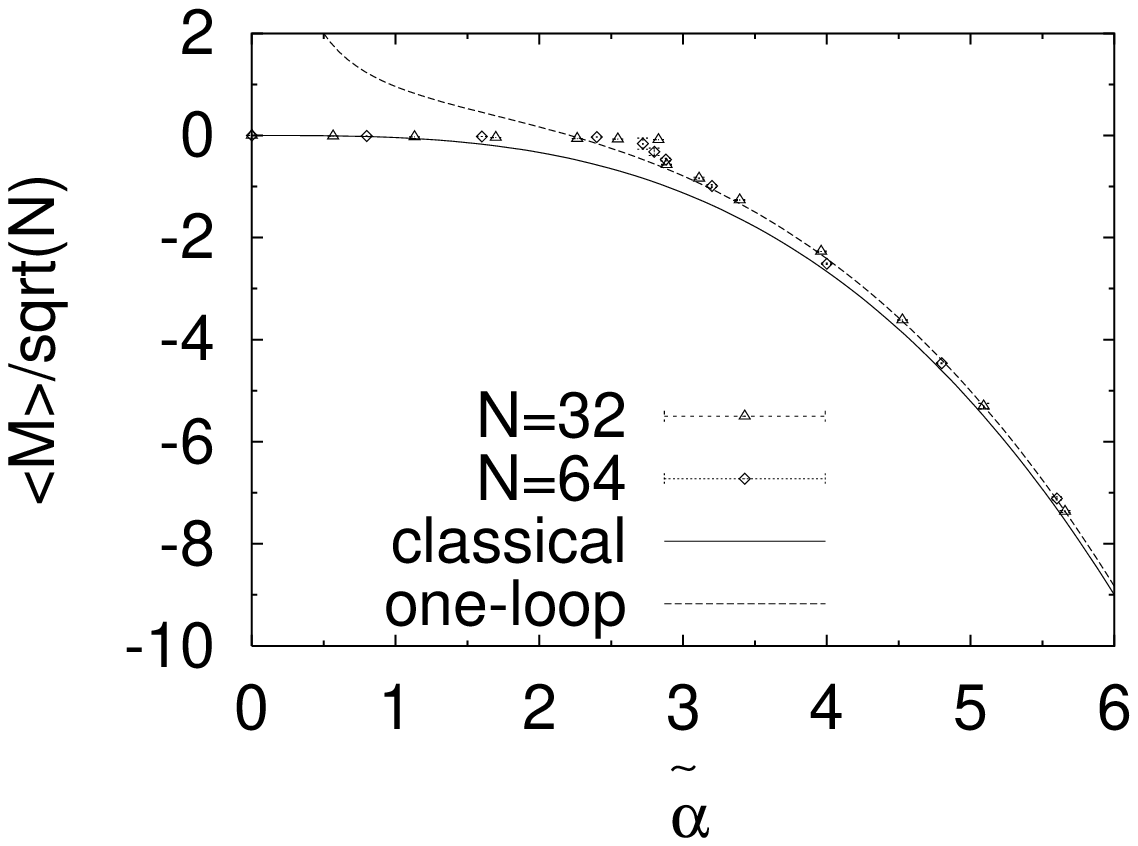,width=7.4cm}
\caption{Various observables are plotted
against $\tilde{\alpha}$ for $N=32,64$.
The initial configuration is taken to be
the $k=2$ coincident fuzzy spheres (\ref{kini}),
and the measurement has been made before they decay into
other states.
}
\label{miscFSk}}

Fig.\ \ref{miscFSk} shows the results for the observables
obtained from 
simulations starting from 
the $k=2$ coincident fuzzy spheres (\ref{kini}).
As in the case of the single fuzzy sphere ($k=1$), we find that 
all the observables show a discontinuity at some critical point,
but this time the critical point is given by
  \begin{eqnarray}
   {\tilde \alpha}_{\rm cr}^{{\rm (l)}k=2} 
  \sim 2.8 \ . 
\label{criticalpointfsk}
  \end{eqnarray}
Above the critical point the results agree very well with the 
one-loop calculation.
This suggests that the one-loop dominance holds for
$k$ coincident fuzzy spheres as well as for the single fuzzy sphere.

We can also understand the observed value 
(\ref{criticalpointfsk}) for the critical point.
As we discussed in Section \ref{stab},
we can put a lower bound on the critical point by
assuming the one-loop dominance 
for the space-time extent $\langle \frac{1}{N} \tr (A_{\mu})^{2} \rangle$.
For general $k$ we find a lower bound 
${\tilde \alpha}_{\rm cr}^{{\rm (l)}k}  > \sqrt{2k}$,
which is satisfied by (\ref{criticalpointfsk}).
In fact from the analysis which leads to the result
(\ref{pred_eff}) for $k=1$, we obtain
\beq
\tilde{\alpha}_{\rm cr}^{{\rm (l)}k} \simeq
\sqrt{k} \left( \frac{8}{3} \right)^{3/4}
\label{pred_eff_k}
\eeq
at large $N$. For $k=2$ it gives
$ \tilde{\alpha}_{\rm cr}^{{\rm (l)}k=2} \simeq
\sqrt{2} \left( \frac{8}{3}\right)^{3/4} = 2.95$,
which is indeed close to (\ref{criticalpointfsk}).

\subsection{Comparison of the one-loop effective action}
\label{comp_eff}

The instability of the 
$k$ coincident fuzzy spheres, which seems to be responsible for the
finite `life time' obeying the power law, may disappear at larger
$N$ as discussed in the Appendix \ref{instability}.
Therefore by Monte Carlo simulation alone, 
we cannot exclude the possibility that 
the $k$ coincident fuzzy spheres become the true vacuum
of the model at some value of $\tilde{\alpha}$ in the large $N$ limit.
Assuming the one-loop dominance, however,
we may discuss this issue rigorously by using
the one-loop effective action, 
which is derived in the Appendix \ref{eff_derive}.
By comparing (\ref{W_irreducible}) and (\ref{W_reducible})
we find that the $k$ coincident fuzzy spheres
have a smaller effective action than the single fuzzy sphere for
\beq
\tilde{\alpha} < 
\left\{ \frac{24 \, \log k}{1-\frac{1}{k^2}}
\right\}^{1/4}  \equiv \tilde{\alpha}_k \ .
\eeq
The critical value $\tilde{\alpha}_k$
is 2.17 for $k=2$, and it grows very slowly with $k$.

Here the existence of the `lower critical point' (\ref{pred_eff_k})
for the $k$ coincident fuzzy spheres
has a dramatic consequence.
Note that the `lower critical point'
$\tilde{\alpha}_{\rm cr}^{{\rm (l)}k} \simeq 2.1 \sqrt{k}$
grows much faster with $k$ than $\tilde{\alpha}_k$.
As a result it turns out that $\tilde{\alpha}_k < 
\tilde{\alpha}_{\rm cr}^{{\rm (l)}k}$ for any $k\ge 2$.
This means that
all the $k$($\ge 2$) coincident fuzzy spheres are actually
taken over by the Yang-Mills phase
at $\tilde \alpha$ where they could have a smaller effective action
than the single fuzzy sphere.
%
%
Thus we conclude that the $k$ coincident fuzzy spheres cannot
be realized as the true vacuum in this model 
{\em even in the large} $N$ {\em limit}.

\section{Summary and discussions} 
\label{section:conclusion}

In this paper we have studied nonperturbative properties
of the matrix model with the Chern-Simons term.
In particular we have demonstrated that the single fuzzy sphere 
emerges dynamically for sufficiently large $\alpha$,
and it describes the true vacuum of the model in that regime.
The model may therefore serve as a nonperturbative definition of 
a NC gauge theory on the fuzzy sphere. It would be interesting
to study various correlation functions from the field theoretical
point of view and to study the UV/IR mixing effects arising from 
noncommutative geometry as has been done for the NC torus 
\cite{Bietenholz:2002ch,Bietenholz:2002ev,Ambjorn:2002nj}.

The dynamical generation of the fuzzy sphere itself is an interesting
phenomenon, since it provides us with a concrete example in which
the space-time that appears dynamically has lower dimensionality
than the original dimensionality that the model can actually describe.
Although the mechanism may be different, our result gives more
plausibility to the scenario for the dynamical generation of the
four-dimensional space-time in the IKKT matrix model
\cite{Aoki:1998vn,NV,sign,Nishimura:2001sx}.

When $\alpha$ is smaller than a critical value, the single fuzzy sphere 
is no more stable, and the large $N$ behavior of the system is 
qualitatively the same as in the pure Yang-Mills model ($\alpha=0$).
Here the space-time looks more like a solid ball with higher density
toward the center due to the one-loop attractive potential.
It would be interesting to study the supersymmetric
case where the one-loop attractive potential is canceled by the fermionic
contributions at the leading order of $N$.
The density distribution actually has an empty region in the center,
which can be explained by the `uncertainty principle', 
but this picture
may depend on how one defines the density distribution when the space-time
is noncommutative. 

The fuzzy sphere phase and the Yang-Mills phase
are separated by a first order phase transition
and we observe a strong hysteresis.
One of the interesting features of the phase transition
is that the lower critical point
behaves as ${\alpha}_{\rm cr} ^{\rm (l)} \sim \frac{2.1}{\sqrt{N}}$,
whereas the upper critical point
behaves as ${\alpha}_{\rm cr} ^{\rm (u)} \sim 0.66$ at large $N$.
The position of the lower critical point 
can be reproduced by considering the effective potential for the
scalar mode on the fuzzy sphere \cite{private}.

Various multi fuzzy spheres appear as meta-stable states 
at sufficiently large $\alpha$,
but we cannot exclude the possibility
that they stabilize eventually in the large $N$ limit.
We have argued, however, that the $k$ coincident fuzzy spheres
cannot be the true vacuum even in the large $N$ limit.
This conclusion was obtained from the comparison of the one-loop
effective action taking account of 
the existence of the `lower critical point' for each $k$.


The quantum corrections around the fuzzy-sphere solutions 
are found to be dominated by the one-loop contribution at large $N$ 
despite the absence of supersymmetry.
This may be understood by the naive power counting argument 
as the one in Ref.\ \cite{0307007}.
It is noteworthy that the `one-loop dominance'
seems to hold even for meta-stable states such as 
the $k$ coincident fuzzy spheres.
Here the measurements in simulations are made before
the meta-stable states actually decay,
while the perturbative calculations are performed omitting the zero modes.

Thanks to the meta-stability of the $k$ coincident fuzzy spheres,
the model may serve also as a toy model
for the dynamical generation of gauge group in matrix models.
In Ref.\ \cite{9903217} it was argued that
the U($k$) gauge symmetry appears in the low-energy effective theory
if the eigenvalues of $A_\mu$ form a cluster of size $k$.
It would be interesting to examine whether such a clustering 
is really taking place in the present model 
at the fully nonperturbative level.
From that point of view, it would be also interesting to search
for a model in which the gauge group of higher rank is generated
in the {\em true vacuum}.

We hope that the present work demonstrates the usefulness of 
Monte Carlo simulations in revealing various interesting dynamics
of matrix models associated in particular with the eigenvalues of the
matrices. The actual dynamics may of course depend much on the model,
and therefore it is important to extend these studies to other models.
Of particular interest is the impact of supersymmetry.
The supersymmetric extension 
of the model we studied in this paper is known \cite{0101102},
but the partition function diverges \cite{0310170}
except for the particular case of $N=2$ \cite{0309264}.
A sensible starting point may therefore be to 
add the three-dimensional Chern-Simons term 
(as the one in the present model)
to the $D=4$ supersymmetric matrix model studied in Ref.\ \cite{0003208}.
Another direction is to study matrix models which have higher dimensional
fuzzy spheres as classical solutions 
\cite{0204256,0209057,Imai:2003ja}.
The stability issue in this case
has not been studied yet due to technical complications,
but the application of our method is straightforward.
Extensions to various homogeneous spaces by replacing the epsilon tensor
in the Chern-Simons term by the structure constant of more general 
Lie algebra may also be of interest.
Finally we add that studying matrix models with a cubic term
may be of relevance to supermatrix models based on super Lie algebra
\cite{supermatrix,0209057}
as well as to the quantum Hall systems \cite{QHE}.



\acknowledgments
We would like to thank Maxime Bagnoud, So Matsuura and Yastoshi Takayama
for their participation at the earlier stage of this work.
We are also grateful to Satoshi Iso, Hikaru Kawai, Noboru Kawamoto,  
Yusuke Kimura, Yoshihisa Kitazawa, Yoshinobu Kuramashi, Xavier Martin,
Denjoe O'Connor, Tetsuya Onogi, Dan Tomino, Badis Ydri and Kentaroh Yoshida 
for helpful discussions. 
The work of T.A., S.B.\ and J.N.\ 
is supported in part by Grant-in-Aid for 
Scientific Research (Nos.\ 01282, P02040 and 14740163, respectively)
from the Ministry of Education, Culture, Sports, Science and Technology. 
T.A.\ and K.N.\ are greatly benefited from discussions at 
"YONUPA Summer School 2003" (YITP-W-03-09) and
``YITP School on Lattice Field Theory'' (YITP-W-02-15)
supported by the Yukawa Institute for Theoretical Physics (YITP).
Part of the computations were carried out on the supercomputers at YITP.


\bigskip 

\appendix
\section{The heat bath algorithm for Monte Carlo simulations}
  \label{heat-bath}
  In this Section we comment on the algorithm used for our Monte Carlo
  simulation. The crucial point is that 
the Chern-Simons term is {\it linear} with respect
to each of $A_{1}$, $A_{2}$ and $A_{3}$.
This allows us to extend the heat bath algorithm 
used for the pure Yang-Mills model ($\alpha = 0$)
in Ref.\ \cite{9811220} to the present model in a straightforward
manner.

Following Ref.\ \cite{9811220} we consider the action
  \beq
  S' = \frac{N}{2} \sum_{\mu < \nu} \tr \left( Q_{\mu \nu}^{2} - 2
  Q_{\mu \nu} G_{\mu\nu} \right) + 2N \sum_{\mu < \nu} \tr
  (A_{\mu}^{2} A_{\nu}^{2})   
  + \frac{2i \alpha N}{3} \epsilon_{\mu \nu \rho} \tr A_{\mu} A_{\nu}
  A_{\rho} \ ,
\label{MCaction}
  \eeq
where $G_{\mu\nu} = \{ A_{\mu}, A_{\nu} \}$.
We have introduced the auxiliary fields 
$Q_{\mu \nu}$ ($1 \leq \mu < \nu \leq 3$),
which are $N \times N$ hermitian matrices.
For $\mu > \nu$ we define $Q_{\mu \nu}$ as $Q_{\mu \nu} = Q_{\nu \mu}$.
Integrating out the auxiliary fields $Q_{\mu \nu}$, we retrieve the
original action (\ref{verydefinition}).

The auxiliary fields $(Q_{\mu  \nu})_{ij}$ can be updated by 
generating a hermitian matrix with an appropriate Gaussian weight 
and shifting it by $G_{\mu\nu}$.
In order to consider how to update $A_{\rho}$,
we extract the part of the action
(\ref{MCaction}) depending on $A_{\rho}$ as
  \begin{eqnarray}
  S' =   2N \, \tr \{ S_{\rho} (A_{\rho})^{2} \}
        -N \, \tr (T_{\rho} A_{\rho}) +  \cdots \ ,
  \end{eqnarray} 
where $S_{\rho}$ and $T_{\rho}$ 
are hermitian matrices defined by
\begin{eqnarray}
   S_{\rho} &=& \sum_{\mu \neq \rho} (A_{\mu})^{2},
   \label{s-update} \\ 
   T_{\rho} &=& \sum_{\mu \neq \rho} (A_{\mu} Q_{\rho \mu}
          + Q_{\rho \mu} A_{\mu})
          - 2\, i \, \alpha \sum_{\mu,\nu}
       \epsilon_{\rho \mu \nu} A_{\mu}A_{\nu} \ . 
\label{t-update}  
  \end{eqnarray}
Since $(A_{\rho})_{ij}$ couples only with $(A_{\rho})_{jk}$, 
where the index $j$ is common, we can update $\frac{N}{2}$ 
components of $A_{\rho}$, which do not have common indices, 
simultaneously.
The only modification needed to incorporate the Chern-Simons term
is the second term in (\ref{t-update}).

\section{The one-loop effective action}
\label{eff_derive}
In this Section we formulate the perturbation theory
and derive the one-loop effective action
for the single fuzzy sphere and the $k$ coincident fuzzy spheres,
which is discussed in Section \ref{comp_eff}.


We decompose $A_{\mu}$ into the classical background $X_{\mu}$ 
and the fluctuation $\tilde A_{\mu}$ as 
\begin{equation}
A_{\mu} = X_{\mu} + {\tilde A}_{\mu} \ ,
\end{equation}
and integrate over $\tilde A_{\mu}$ perturbatively.
In order to remove the zero modes associated with the SU($N$)
invariance, we introduce the gauge fixing term and the
corresponding ghost term
\begin{eqnarray}
S_{\rm g.f.} &=& -\frac{N}{2}\text{tr}[X_\mu, A_\mu]^2 \ , \\
S_{\rm ghost}&=& -N\text{tr} \left([X_\mu,\bar{c}][A_\mu,c] \right) \ ,
\end{eqnarray}
where $c$ and $\bar c$ are the ghost and anti-ghost fields, respectively. 
The total action reads
\beq
S_{\rm total} = S + S_{\rm g.f.} + S_{\rm ghost} \label{totalaction} \ ,
\eeq
which is given explicitly as
\beqa
S_{\rm total} &=& S[X]+ S_{\rm kin} + S_{\rm int}  \ ,\\
S_{\rm kin}&=&\frac{1}{2}\,  N \, \tr \left( 
\tilde A_\mu [X_\lambda , [X_\lambda , \tilde A_\mu ]] \right)
+ N \, \tr \Bigl( \bar c \, [X_\lambda , [X_\lambda , c ]] \Bigr)   \non
&~& - N \, \tr \Bigl\{
\Bigl( [X_\mu , X_\nu] - i \alpha \epsilon_{\mu\nu\rho} X_\rho \Bigr)
[\tilde A_\mu , \tilde A_\nu ]\Bigr\} \ ,
\label{SUalg_term}
\\
S_{\rm int}&=&-N \, \tr \left( 
[\tilde A_{\mu} , \tilde A_{\nu} ][X_{\mu} ,\tilde A_{\nu} ] \right) 
- \frac{1}{4}\,  N\, 
\tr \left( [\tilde A_{\mu} , \tilde A_{\nu} ]^2 \right) \non
&&+ \frac{2}{3}\, i \, \alpha \, \epsilon_{\mu \nu \rho} \, N \,
 \tr \left( \tilde A_{\mu}  \tilde A_{\nu}  \tilde A_{\rho}  \right)
+ N \, \tr \left( \bar c \, [X_{\mu}, [\tilde A_{\mu} , c]] \right) \ .
\eeqa
The linear terms in $X_\mu$ cancel since $X_\mu$ is assumed to
satisfy the classical equation of motion (\ref{eom}).
Here and henceforth 
we restrict ourselves to the case where the classical solution $X_\mu$
is proportional to $\alpha$.
Then, by rescaling the matrices as $A_\mu \mapsto \alpha \, A_\mu$,
$c \mapsto \alpha \, c$, $\bar c \mapsto \alpha \, \bar c$, 
all the terms in the total action $S_{\rm total}$ will be proportional
to $\alpha ^{4}$.
This means that the expansion parameter of the present perturbation
theory is $\frac{1}{\alpha ^{4}}$.

The effective action $W$ is defined by
\beq
\ee ^{-W} = \int  \dd \tilde A \, \dd c 
\, \dd \bar c \, \ee ^{-S_{\rm total}}  \ ,
\eeq
and we calculate it as a perturbative expansion 
$W = \sum_{j=0}^{\infty} W_j$,
where $W_j = \mbox{O}(\alpha^{4(1-j)})$.
The classical part is given by
$W_0 = S[X]$, which is nothing but the action (\ref{verydefinition})
evaluated at the classical solution $A_\mu = X_\mu$.

In this Section we consider the solutions of the form (\ref{fssolution}),
for which the third term in (\ref{SUalg_term}) vanishes.
\footnote{
We will consider a more general case in Appendix
\ref{instability}.
}
Then the kinetic term (\ref{SUalg_term}) can be written as
\beq
S_{\rm kin} =
N \alpha^2  \tr\left[
\frac{1}{2}  \, \tilde A_\mu  ({\cal P}_\lambda )^2  \tilde A_\mu  
 +  \bar c \, ({\cal P}_\lambda )^2  c \right] \ ,
\label{SQ_fs}  
\eeq
where we have introduced the operator ${\cal P}_{\mu}$
\begin{equation}
{\cal P}_{\mu} M \equiv [X_{\mu}, M]  \ ,
\end{equation}
which acts on the space of $N \times N$ matrices.
The one-loop term can be obtained as
\beq
W_1 = 
\frac{1}{2} \,  {\cal T}r \log \left\{ N ({\cal P}_\mu)^2 \right \}  
\ , \label{one-loop_Seff}
\eeq
where the symbol ${\cal T}r$ denotes
the trace of such an operator.
%

\subsection{The single fuzzy sphere}

Let us first consider the single fuzzy sphere
$X_{\mu} = \alpha L_\mu ^{(N)}$.
The classical part is given by
\begin{equation}
W_0 = -\frac{1}{24}\alpha^4 N^2 (N^2 -1) \ .
\label{Scl_IR}
\end{equation}
In order to calculate the one-loop term, 
we note that the eigenvalue problem
of the operator $({\cal P}_\lambda)^2$ in the present case
can be solved by the matrix analog of the spherical harmonics
$Y_{lm}$ ($0 \le l \le N-1$, $-l \le m \le l$), 
which form a complete basis in the space of $N \times N$ matrices
and have the property
\beq
({\cal P}_\mu)^2 Y_{lm} = \alpha^2 \, l \, (l+1) \, Y_{lm} \ .
\label{eigenY}
\eeq
Similarly to the usual spherical harmonics,
they satisfy
\beqa
\frac{1}{N}\tr \left(  Y_{lm}^\dag  Y_{l'm'}  \right)
&=& \delta_{ll'}\delta_{mm'} \ ,
\label{Yortho} \\
Y_{lm}^\dag &=& (-1)^m Y_{l,-m} \ .
\label{Yconjg}
\eeqa
See Ref.\ \cite{0101102} for other properties.
From (\ref{eigenY}) we find that $Y_{00}= {\bf 1}_N$ is a zero
mode, which corresponds to the invariance of the action 
(\ref{verydefinition})
under
$A_\mu \mapsto A_\mu + \alpha_\mu {\bf 1}_N$.
However, we should omit this mode when we take the trace
${\cal T}r$ in (\ref{one-loop_Seff}) since all the matrices
$\tilde{A}_\mu$, $c$ and $\bar c$ are supposed to be traceless.
Thus we obtain the one-loop contribution as
\beq
W_1 = \frac{1}{2}\sum_{l=1}^{N-1} 
(2l+1) \log \left[ N \, \alpha^2 \,  l(l+1) \right]  \ .
\eeq
At large $N$ the one-loop effective action is obtained as
\beq
W_{\rm 1-loop} \simeq N^2 \left( -\frac{\tilde{\alpha}^4}{24}
+ \log \tilde{\alpha} +\log N \right)  \ ,
\label{W_irreducible} 
\eeq
where $\tilde{\alpha} = \alpha \sqrt{N} $.

\subsection{The $k$ coincident fuzzy spheres}
\label{coin_effact}

Next we consider the $k$ coincident fuzzy spheres 
\beq
X_{\mu} = \alpha L_{\mu}^{(n)} \otimes {\bf 1}_{k} \ .
\label{k_coinX}
\eeq
By setting $n_1 = \cdots = n_k \equiv n \left( =  \frac{N}{k} \right)$
in (\ref{multiaction}), the classical part is obtained as
\beq
W_0 = - \frac{\alpha^{4} N^2}{24}  (n^2 - 1)  \ .
\label{multi_cl_action}
\eeq
In order to solve the eigenvalue problem of the operator 
$({\cal P}_\lambda)^2$ in the present case,
we consider the $n \times n$ version of the matrix spherical harmonics
$Y'_{lm}$ and introduce
a $k\times k$ matrix ${\bf e}^{(a,b)}$,
whose ($a,b$) element is 1 and all the other elements are zero.
Then as a complete basis of $N \times N$ matrices,
we define
\beq
Y_{lm}^{(a,b)} \equiv Y'_{lm} \otimes {\bf e}^{(a,b)} \ ,
\eeq
which has the property
\beq
({\cal P}_\lambda)^2 Y_{lm}^{(a,b)}
 = \alpha^2 \, l \, (l+1) \, Y_{lm}^{(a,b)} \ .
\eeq
Here $Y_{00}^{(a,b)}$ for all the $(a,b)$ blocks are the zero modes.
The trace part $\sum_{a=1}^{k} Y_{00}^{(a,a)} = {\bf 1}_N $
should be omitted as before, but not the others.
We will study the remaining $(k^2 - 1)$
zero modes in Appendix \ref{instability},
but here let us consider only the non-zero modes.
Then the one-loop contribution $W_1$ is obtained as
\beq
W_1 =
\frac{1}{2} k^2 \sum_{l=1}^{n-1} 
(2l+1) \log \left[ N \alpha^2 l(l+1)  \right]  \ .
\label{coin_ol}
\eeq
At large $N$ the one-loop effective action (neglecting the zero modes)
reads
\beq
W_{\rm 1-loop} \simeq N^2 \left( -\frac{\tilde{\alpha}^4}{24 \, k^2}
+ \log \tilde{\alpha} + \log\frac{N}{k}  \right)  \ .
\label{W_reducible}
\eeq




\section{The one-loop calculation of various observables}
\label{one-loop-obs-appendix}

In this Section we apply 
the perturbation theory discussed in Appendix \ref{eff_derive}
to the one-loop calculation of various observables
which are studied by Monte Carlo simulations in this paper.
Here we take the background to be $k$ coincident fuzzy spheres 
(\ref{k_coinX}), but the results for the single fuzzy sphere
can be readily obtained by setting $k=1$.
As in Appendix \ref{coin_effact}, we omit the zero modes 
for $k \ge 2$, which will be discussed in Appendix \ref{instability}.

We note that 
the number of loops in the relevant diagrams
can be less than the order of 
$\frac{1}{\alpha^4}$ in the perturbative expansion
since we are expanding the theory around a nontrivial background.
At the one-loop level, the only nontrivial task is to 
evaluate the tadpole $\langle (\tilde A_\mu)_{ij} \rangle$ 
explicitly.



%



\subsection{Propagators and the tadpole}
Using the properties (\ref{Yortho}), (\ref{Yconjg}) of the 
matrix spherical harmonics,
the propagators for $\tilde A_\mu$ and the ghosts
are given as
\beqa
\left\langle (\tilde A_\mu)_{ij} (\tilde A_\nu)_{kl}
\right\rangle _0
&=& 
\delta_{\mu\nu} \frac{1}{n}
\sum_{ab} \sum_{l=1}^{n-1} \sum_{m=-l}^l 
\frac{(-1)^{m}}{N \alpha^2 l(l +1)}
\left( Y_{lm}^{(a,b)}  \right)_{ij}
\left( Y_{l, -m}^{(b,a)} \right)_{kl} \ , 
\label{prop_Amat}  \\
\Bigl \langle(c)_{ij} (\bar c)_{kl} \Bigr\rangle_0 &=& 
\frac{1}{n} \sum_{ab} \sum_{l=1}^{n-1} \sum_{m=-l}^l 
\frac{(-1)^{m}}{N \alpha^2 l (l +1)}
\left( Y_{lm}^{(a,b)}  \right)_{ij}
\left( Y_{l, -m}^{(b,a)} \right)_{kl} \ ,
\label{prop_cmat}
\eeqa
where the symbol $\langle \ \cdot \ \rangle_0$ refers to the
expectation value calculated using the kinetic term $S_{\rm kin}$ 
in (\ref{SUalg_term}) only.


Due to the symmetries, the tadpole
$\langle \tilde A_\mu \rangle$ 
can be expressed as 
\beq
\langle\tilde A_\mu \rangle = c \,  X_\mu 
\eeq
with some coefficient $c$. Using the identity
\begin{eqnarray}
\tr \left( X_\mu \langle \tilde A_\mu \rangle \right) &=& 
c \, \tr (X_\mu X_\mu) \n \\
&=& c  \, \frac{\alpha^2}{4} (n^2 -1 ) N \ ,
\label{inner-product}
\end{eqnarray}
the coefficient $c$ can be determined by
calculating the left hand side of (\ref{inner-product}).

At the leading order in $\frac{1}{\alpha^4}$,
we have
\begin{eqnarray}
\frac{1}{N} \tr
\left( X_\mu \langle \tilde A_\mu \rangle_{\rm 1-loop} \right)
&=& 
 \left\langle \tr(X_\mu \tilde A_\mu ) \, 
\tr \left([\tilde A_{\nu} ,\tilde A_{\rho} ][X_{\nu} ,\tilde
A_{\rho} ] \right) \right\rangle_0 \n \\
&~& 
- \left\langle \tr(X_\mu \tilde A_\mu)
 \,  \tr\left( \frac{2}{3}\, i \, \alpha \, 
\epsilon_{\nu \rho \sigma} \tilde A_{\nu}  \tilde A_{\rho}  \tilde
A_{\sigma}  \right)\right\rangle_0 \n \\
&~&  -  \left\langle \tr(X_\mu \tilde A_\mu)  \, \tr
\left( \bar c \, [X_{\nu} ,[\tilde A_{\nu} ,c] ] \right) 
\right\rangle_0 \ .
\label{tadpole}
\end{eqnarray}
Using the fact that $X_\mu$ is a linear combination of 
$(Y'_{l=1,m} \otimes {\bf 1}_{k})$, we 
can calculate (\ref{tadpole}) in a straightforward manner.
After some algebra we arrive at
\begin{eqnarray}
\tr \left( X_\mu \langle \tilde A_\mu \rangle_{\rm 1-loop} \right)=
-\frac{k^2}{2 N \alpha^2}
(n^2 -1) \ .
\label{tr_L^red_A}
\end{eqnarray}
Using (\ref{inner-product}) we obtain
\begin{equation} 
\langle\tilde A_\mu \rangle_{\rm 1-loop} = 
-\frac{2k^2}{N^2 \alpha^4}  \,  X_\mu  \ .
\end{equation}


\subsection{One-loop results for various observables}
Using the propagator and the tadpole obtained in the previous 
Section, we can evaluate various observables easily at the one-loop
level.

The two-point function
$\langle \frac{1}{N} \tr(A_\mu)^2 \rangle$
can be evaluated as
\beqa
\left\langle \frac{1}{N} \tr (A_\mu)^2 \right\rangle
_{\rm 1-loop}
&=&\frac{1}{N}\left[\tr(X_\mu X_\mu)
+  2 \, \tr \left( X_\mu \langle\tilde A_\mu\rangle _{\rm 1-loop} \right)
+\langle\tr (\tilde A_\mu)^2 \rangle_0 \right] \n \\
&=&
\alpha^2 \, \left[
\frac{1}{4} (n^2 -1)
- \frac{1}{\alpha^4}
\left( 1  - \frac{1}{n^2} \right)
+ \frac{3}{\alpha^4 n^2}
\sum_{l=1}^{n-1} \frac{2 l + 1 }{l(l+1)} \right] \ .
\label{tr_A^2}
\eeqa
At large $N$ with fixed $\tilde \alpha = \sqrt{N} \alpha$, we obtain
\beq
\frac{1}{N}
\left\langle \frac{1}{N} \tr(A_\mu)^2 \right\rangle_{\rm 1-loop}
\simeq \tilde{\alpha}^2 
\left[ \frac{1}{4 \, k^2} -\frac{1}{\tilde{\alpha}^4} \right] \ . 
\label{tr_A^2_largeN}
\eeq
The Chern-Simons term $\langle M \rangle$ can be evaluated as
\begin{eqnarray}
\langle M \rangle _{\rm 1-loop}
&=& \frac{2i}{3N}  \epsilon_{\mu\nu\rho}
\left[  
\tr(X_\mu X_\nu X_\rho)
+3\, \tr \left( X_\mu X_\nu \langle 
\tilde A_\rho \rangle_{\rm 1-loop} \right) 
\right] \n \\
&=&
-\frac{\alpha^3}{6}(n^2-1) + \frac{1}{\alpha}
\left( 1-\frac{1}{n^2} \right) \ .
\label{o2}
\end{eqnarray}
At large $N$ with fixed $\tilde \alpha = \sqrt{N} \alpha$, we get
\beq
\frac{1}{\sqrt{N}}\langle M \rangle _{\rm 1-loop}
\simeq
-\frac{\tilde{\alpha}^3}{6 \, k^2} + \frac{1}{\tilde{\alpha}} \ . 
\label{o2_largeN}
\eeq
The observable 
$\langle \frac{1}{N} \tr F^2 \rangle$
can be calculated in a similar manner, but 
we can also deduce it from the exact result (\ref{sde-quantity}) as
\begin{equation}
\left \langle \frac{1}{N} \tr 
(F_{\mu\nu})^2 \right \rangle _{\rm 1-loop}
= 3\left( 1-\frac{1}{N^{2}} \right)
- 3 \alpha \langle M \rangle _{\rm 1-loop}
\simeq \frac{\tilde{\alpha}^4}{2 \, k^2} \ . 
\label{o1}
\end{equation}
Combining (\ref{o2}) and (\ref{o1}), 
we get 
\beq
\frac{1}{N^2} \langle S \rangle _{\rm 1-loop}
%
\simeq -\frac{\tilde{\alpha}^4}{24 \, k^2} + 1 \ . 
\label{appendix_(S)}
\eeq

Next we calculate the variance $\sigma^2$ of the eigenvalue distribution
$f(x)$ of the Casimir operator studied in Section \ref{section:width}.
The first term in (\ref{sigma2_def})
can be evaluated as follows.
\begin{eqnarray}
\frac{1}{N} \left \langle \tr
\left( {A_\mu}^2 \right)^2  \right\rangle_{\rm 1-loop} &=&
\frac{1}{N}\left[ 
\tr \left( {X_\mu}^2 {X_\nu}^2 \right) 
+2\tr\left({X_\mu}^2 \left\{X_\nu,
\langle \tilde A_\nu \rangle_{\rm 1-loop} 
\right\}\right) \right. \n \\
&& + \tr \left( \left\{ {X_\mu}^2, 
\left\langle {\tilde {A_\nu}}^2 \right \rangle_0 
\right\} \right)
+2\tr(X_\nu X_\mu \langle \tilde A_\mu \tilde 
A_\nu \rangle_0  ) 
\n \\
&& \left. 
+ \langle \tr(X_\mu \tilde A_\mu X_\nu \tilde A_\nu) 
\rangle _0 
+ \langle \tr(\tilde A_\mu X_\mu \tilde A_\nu X_\nu) 
\rangle_0  \right] \n \\
&=&  \frac{(n^2-1)^2}{16} \alpha^4 -\frac{(n^2 -1)^2}{2n^2}  \n \\
&~& +\left( 1-\frac{1}{n^2} \right)
\left[ \frac{5}{2}\sum_{l=1}^{n-1}\frac{2l+1}{l(l+1)} -1 \right] .
\label{tr(a2)2}
\end{eqnarray}
The second term in (\ref{sigma2_def}) is given by
the square of (\ref{tr_A^2}).
The classical part of $\sigma^2$ cancels exactly as it
should. Note that since the leading O($N^2$) terms
in the one-loop part also cancel, we have to keep track of the
subleading O($\log N$) terms.
Using the asymptotic behavior
\beq
\sum_{l=1}^{n-1}\frac{2l+1}{l(l+1)} \simeq 2 \, \log n 
\eeq
at large $n$, we obtain
\beq
(\sigma^2)_{\rm 1-loop} \simeq 2 \, \log \left( \frac{N}{k}  \right) \ . 
\eeq

\subsection{An alternative derivation}

Since $\tr F^2$ and $M$ are the operators 
that appear in the action $S$, we can obtain their expectation values
easily by using the effective action 
calculated for the $k$ coincident fuzzy sphere
in Appendix \ref{coin_effact}.
Let us consider the action
\begin{eqnarray}
   S(\beta_{1}, \beta_{2}, \alpha)
 = N \tr \left( - \frac{\beta_{1}}{4} [A_{\mu},
   A_{\nu}]^{2} + \frac{2i \alpha \beta_{2}}{3}  
   \epsilon_{\mu \nu \rho} A_{\mu} A_{\nu} A_{\rho}
   \right) \ , \label{verydefinition2}
\end{eqnarray}
where we have introduced two free parameters
$\beta_{1}$ and $\beta_{2}$,
and define the corresponding effective action by
\begin{eqnarray}
  \ee ^ {- W(\beta_{1}, \beta_{2}, \alpha)} 
 =  \int \dd A \, \ee ^{-S(\beta_{1}, \beta_{2}, \alpha)} \ .
\label{oneloopefb12}
\end{eqnarray}
Then $\langle \tr (F_{\mu\nu})^2 \rangle$ 
and $\langle M \rangle$ can be obtained by
\beqa
\left\langle \frac{1}{N} \tr (F_{\mu \nu})^2 \right\rangle &=& 
\frac{4}{N^{2}} \left. \frac{\partial W}{\partial \beta_{1}} 
\right|_{\beta_{1} = \beta_{2} = 1}  \ ,
\label{one-loopf-sq2}  \\
\langle M \rangle &=& \frac{1}{\alpha N^{2}} 
\left. \frac{\partial W}{\partial \beta_{2}} 
 \right|_{\beta_{1} = \beta_{2} = 1} \ .
\label{one-loopcs-a2} 
\end{eqnarray}
By rescaling the integration variables as
$A_{\mu} \mapsto \beta_{1}^{- \frac{1}{4}}  A_{\mu}$,
we find
\beq
   W(\beta_{1}, \beta_{2}, \alpha)
= \frac{3}{4}(N^{2}-1) \log \beta_{1} + 
W(1,1,\alpha \, \beta_{1}^{-\frac{3}{4}} \beta_{2} )   \ .
\eeq
Using the one-loop result
\beq
W(1,1,\alpha)_{\rm 1-loop} =
 - \frac{\alpha^{4} N^2}{24}  (n^2 - 1) 
+ \frac{1}{2} k^2 \sum_{l=1}^{n-1} 
(2l+1) \log \left[ N \alpha^2 l(l+1)  \right]  \ ,
\eeq
which follows from (\ref{multi_cl_action}) and (\ref{coin_ol}),
we can reproduce (\ref{o2}) and (\ref{o1}).



\section{Instability of the $k$ coincident fuzzy spheres}
\label{instability}

In the Appendix \ref{coin_effact}
we encountered zero modes in the one-loop
perturbative expansion around $k$ coincident fuzzy spheres.
This corresponds to the fact that the action (\ref{verydefinition})
does not change up
to the second order perturbation by the deformation
\beq
A_\mu = 
\alpha L_{\mu}^{(n)} \otimes {\bf 1}_{k}
+  {\bf 1}_{n} \otimes \delta H_\mu \ ,
\label{configH}
\eeq
where $\delta H_\mu$ ($\mu = 1, 2, 3$) 
are $k \times k$ hermitian matrices.
In fact the change of the action is given by
\beq
\delta S =  N
\, n \, \tr_k \left( - \frac{1}{4} 
\, [\delta H_{\mu}, \delta H_{\nu}]^{2} + 
\frac{2}{3}\, i \, \alpha 
\, \epsilon_{\mu \nu \rho} 
\delta H_{\mu} \delta H_{\nu} \delta H_{\rho}
   \right) , 
\label{deltaS} 
\eeq
where the symbol $\tr_k$ refers to the trace 
of a $k\times k$ matrix.
Note that $\delta S$ can always be made negative 
for small $\delta H_\mu$ by choosing its sign appropriately 
when $[\delta H_{\mu} , \delta H_{\nu}] \neq 0$.
Thus the $k$ coincident fuzzy spheres
are actually not a local minimum of the {\em classical} action,
although this does not exclude the possibility that they
stabilize due to {\em quantum effects}
at finite $\tilde{\alpha}$ in the large $N$ limit.


When $\delta H_\mu$ ($\mu = 1, 2, 3$) commute with each other,
$\delta S$ in (\ref{deltaS}) vanishes, which means that such a
deformation gives rise to a flat direction.
In fact this is due to the fact that 
the configuration (\ref{configH})
in this case satisfies the equation of motion (\ref{eom})
for finite $\delta H_\mu$. Let us consider whether the
$k$ coincident fuzzy spheres are stable against such a deformation
quantum mechanically.
For that purpose we calculate the one-loop effective action
around the deformed configuration.
Note first that 
we can diagonalize $\delta H_\mu$
by applying an SU($N$) transformation
$A_\mu \mapsto U A_\mu U^\dag$, where $U$ is of the form
$U =  {\bf 1}_{n} \otimes V$ with $V \in \mbox{SU}(k)$.
We may therefore consider a classical solution
\beq
X_\mu = 
\alpha \, \Bigl( L_{\mu}^{(n)} \otimes {\bf 1}_{k}
+  {\bf 1}_{n} \otimes  C_\mu \Bigr) \ ,
\label{XwithC}
\eeq
where
\beq
C_\mu = \mbox{diag} (c_\mu ^{(1)} , \cdots , c_\mu ^{(k)} ) 
\eeq
without loss of generality.
In order to make the configuration $X_\mu$ traceless, 
the $k\times k$ matrices $C_\mu$ should also be traceless.
\begin{equation}   
\tr_k \, \left(C_\mu \right) = \sum_{a=1}^{k} c_\mu^{(a)} = 0 \ .
\label{tracelessC}
\end{equation}
The second term in (\ref{XwithC}) corresponds to
shifting each of the $k$ coincident fuzzy spheres,
where the three-dimensional shift vector 
for the $a$-th sphere is given by 
$\alpha \, c_\mu ^{(a)}$ ($a=1, \cdots , k$).

Let us therefore calculate the one-loop effective action
around the classical solution (\ref{XwithC})
as we did for the $k$ coincident fuzzy spheres
in Appendix \ref{eff_derive}.
The classical part of the effective action is the same as 
the $C_\mu = 0$ case (\ref{multi_cl_action}).
At the one-loop level,
the shifted configuration (\ref{XwithC}) has the same number of 
zero modes as the $k$ coincident fuzzy spheres.
Among them, $(k-1)$ corresponds to changing 
$c_\mu ^{(a)}$ ($a=1, \cdots , k$) 
under the constraint (\ref{tracelessC}), so we do not have to integrate
over them in the present context.
The remaining $k(k-1)$ modes correspond to the `noncommutative shifts'
(\ref{configH}) discussed above, but we simply omit them as we did 
in the $C_\mu = 0$ case.

The kinetic term (\ref{SUalg_term}) in the present case is given by
\beq
S_{\rm kin} =
N \alpha^2  \tr\left[
\frac{1}{2}  \, \tilde A_\mu 
\left\{ ({\cal L}_\lambda + {\cal C}_{\lambda})^2 \, \delta_{\mu \nu} 
+ 2 \, i  \, \epsilon_{\mu \lambda \nu} \, {\cal C}_\lambda
\right\} \tilde A_\nu  
 +  \bar c 
({\cal L}_\lambda + {\cal C}_{\lambda})^2  c \right] \ ,
\label{SQ}  
\eeq
where we introduced the operators ${\cal L}_\mu$ and ${\cal C}_\mu$ as
\beqa
{\cal L}_\mu M &=& \Bigl[ (L_\mu ^{(n)} \otimes {\bf 1}_k) , M  \Bigr] \ ,
 \\
{\cal C}_\mu M &=& \Bigl[ ({\bf 1}_n \otimes C_\mu) , M \Bigr] \ ,
\eeqa
which act on the space of $N \times N$ matrices.
The one-loop term can be given as
\beqa
W_1 &=&
\frac{1}{2} \, {\cal T}r \, \tr ' \log \Bigl[ N \alpha^2 \left\{ 
({\cal L}_\lambda + {\cal C}_\lambda )^2 \, \delta_{\mu\rho} 
+ 2 \, i \,  \epsilon_{\mu\nu\rho} \, {\cal C}_\nu \right\} 
\Bigr]   \n \\
&~& - {\cal T}r \log \Bigl[ N \alpha^2 \left\{ 
({\cal L}_\lambda + {\cal C}_\lambda )^2 \right\} 
\Bigr]  \ ,
\label{Weff1}
\eeqa
where the trace $\tr '$ is taken over Lorentz indices.
We calculate $W_1$ as an expansion with respect to $C_\mu$
using the formulae
\beqa
{\cal L}_1 Y_{lm}^{(a,b)} &=&
\frac{1}{2}\left( b_{l,m+1} Y_{l,m+1}^{(a,b)} +
b_{l,m} Y_{l,m-1}^{(a,b)} \right) \  , \\
{\cal L}_2 Y_{lm}^{(a,b)} &=&
\frac{1}{2i}\left( b_{l,m+1} Y_{l,m+1}^{(a,b)} -
b_{l,m} Y_{l,m-1}^{(a,b)} \right) \  , \\
{\cal L}_3 Y_{lm}^{(a,b)} &=& m Y_{l,m}^{(a,b)} \  , \\
({\cal L}_\lambda)^2 Y_{lm}^{(a,b)} &=& l(l+1)Y_{l,m}^{(a,b)} \ , \\
{\cal C}_\lambda Y_{lm}^{(a,b)} &=&
c_\lambda^{(a,b)} Y_{lm}^{(a,b)} \ ,  
\end{eqnarray}
where 
$c_\rho^{(a,b)} = c_\rho^{(a)}-c_\rho^{(b)}$
and $b_{l,m}$ is given by  
\beq
b_{l,m}=\sqrt{l(l+1)-m(m-1)} \ .
\eeq
After some algebra, we get
\beq
W_1 =
\frac{1}{2}{\cal T}r \log \Bigl[ N \alpha ^2   
({\cal L}_\lambda)^2 \Bigr] 
+ \kappa \,
\sum_{ab} \left( c_\rho^{(a,b)} \right) ^2 
+ \mbox{O}(C_\rho^3)    \ .
\label{shifted_ol}
\eeq
The first term in (\ref{shifted_ol})
is nothing but the result (\ref{coin_ol})
for the $k$ coincident fuzzy spheres.
The coefficient $\kappa$ is given by
\beq
\kappa 
= \frac{1}{6} \sum_{l=1}^{n-1} \frac{2l+1}{l(l+1)} 
- 2 \left( 1 - \frac{1}{n^2}\right) \ .
\label{kappa_def}
\eeq
Note that the first term in (\ref{kappa_def}) grows as 
$\frac{1}{3}\log n$ at large $n$.
By performing the sum in (\ref{kappa_def}) numerically,
we find that the coefficient $\kappa$ changes its sign 
from negative to positive at $n = 374$.


For the cases studied in Section \ref{life}, where $n=2,4,8$,
the $k$ coincident fuzzy spheres are therefore
unstable for such a shift.
Note, however, that the instability for the commutative shift
is a quantum effect unlike 
the instability for the non-commutative shift.
Since the `life time' obeys the same power law up to huge values
of $\alpha$, we consider that the instability responsible
for the observed power law is dominated by the
non-commutative shift.

For a more general multi-fuzzy-sphere state $\langle 3,5 \rangle$
studied in Section \ref{evo}, on the other hand,
we don't have a counterpart of the noncommutative shift.
\footnote{
In Ref.\ \cite{0108002}
the interaction between two fuzzy spheres with different
radii has been calculated.
}
The observed power law therefore suggests the existence
of {\em quantum} instability for the commutative shift.
Thus the instability for the non-coincident case is expected
to be much weaker than the coincident case.
This is consistent with our observation 
that the power ``3.8'' for the former
is much larger than $\frac{4}{3}$ for the latter.


\end{document}